\documentclass[%
  aip,jcp,
 amsmath,amssymb,
 reprint,%
]{revtex4-1}

\usepackage{graphicx}
\usepackage{epstopdf}
\usepackage{dcolumn}
\usepackage{bm}

\usepackage[utf8]{inputenc}
\usepackage[T1]{fontenc}
\usepackage{mathptmx}

\begin{document}


\title[Perturbation Potentials in Alchemical Calculations]{Perturbation Potentials to Overcome Order/Disorder Transitions in Alchemical Binding Free Energy Calculations}

\author{Rajat K. Pal}
\altaffiliation{Ph.D. Program in Biochemistry, The Graduate Center of the City University of New York, New York, NY}
\affiliation{Department of Chemistry, Brooklyn College of the City University of New York, New York, NY}

\author{Emilio Gallicchio}
\email{egallicchio@brooklyn.cuny.edu}
\altaffiliation{Ph.D. Program in Biochemistry, The Graduate Center of the City University of New York, New York, NY}
\altaffiliation{Ph.D. Program in Chemistry, The Graduate Center of the City University of New York, New York, NY}
\affiliation{Department of Chemistry, Brooklyn College of the City University of New York, New York, NY}

\date{\today}

\begin{abstract}
  We investigate the role of order/disorder transitions in alchemical simulations of protein-ligand absolute binding free energies. We show, in the context of a potential of mean force description, that for a benchmarking system (the complex between the L99A mutant of T4 lysozyme and 3-iodotoluene) and for a more challenging system relevant for medicinal applications (the complex of the farnesoid X receptor and inhibitor 26 from a recent D3R challenge) that order/disorder transitions can significantly hamper Hamiltonian replica exchange sampling efficiency and slow down the rate of equilibration of binding free energy estimates. We further show that our analytical model of alchemical binding combined with the formalism developed by Straub et al.\ for the treatment of order/disorder transitions of molecular systems can be successfully employed to analyze the transitions and help design alchemical schedules and soft-core functions that avoid or reduce the adverse effects of rare binding/unbinding transitions. The results of this work pave the way for the application of these techniques to the alchemical estimation with explicit solvation of hydration free energies and absolute binding free energies of systems undergoing order/disorder transitions.
\end{abstract}

\maketitle

\section{\label{sec:intro}Introduction}

%

The accurate modeling of molecular recognition remains one of the most challenging problem in computational molecular biophysics.\cite{GallicchioSAMPL4,gaieb2018d3r} Molecular simulations of molecular binding are limited by the incomplete description of the chemical system,\cite{Mobley:Dill:review:2009,pal2016SAMPL5,henderson2018ligand,cruzeiro2018redox} by the quality of energy functions\cite{harder2015opls3,Albaugh2016advanced,wang2017building,roos2019opls3e} as well as, and probably to a less understood extent, by insufficient conformational sampling.\cite{Lapelosa2011,Mobley2012,bodnarchuk2014strategies,procacci2019solvation}

One of the primary applications of computational models of molecular binding is to provide an estimate of the standard free energy of binding, $\Delta G_{b}^{\circ}$, or, equivalently, the equilibrium constant, $K_{b}$, for the association equilibrium $R+L\leftrightharpoons RL$, between two molecules $R$ and $L$. For example, the binding of a drug molecule to a protein receptor.\cite{Jorgensen2009} While many strategies have been proposed,\cite{michel2010prediction,Gallicchio2011adv,limongelli2013funnel,deng2018comparing} alchemical relative binding free energy methods\cite{lybrand:mccammon1986} are emerging as the leading approaches in pharmaceuticals applications.\cite{Shirts:Mobley:Chodera:2007:review,Wang2012,homeyer2014binding,abel2017advancing,lee2018gpu,Zou2019chemarxiv} In the latter context, in particular, there have been significant advances in controlling errors related to conformational sampling.\cite{wang2011replica,hauser2018predicting}

We focus here on strategies to improve the estimation of absolute binding free energies by alchemical methods.\cite{Jorgensen:Buckner:Boudon:Rives:88,Gilson:Given:Bush:McCammon:97,Deng2009,Chodera:Mobley:cosb2011,kilburg2018assessment} While more challenging to obtain than relative binding free energies, absolute binding free energies provide a more stringent assessment of protocols and force fields\cite{mobley2017predicting} and are better suited for ranking dissimilar compounds in virtual screening applications,\cite{GallicchioSAMPL4,bfzhang2016rnaseh} as well as for investigating binding specificity.\cite{deng2017resolving}

In this work, we illustrate how, analogously to conventional chemical systems as a function of temperature, alchemical systems can undergo order/disorder phase transitions along the alchemical path, and that these cause entropic bottlenecks which hinder the equilibration and convergence of binding free energy estimates. We analyze this phenomenon using the analytic theory of alchemical binding we recently proposed\cite{kilburg2018analytical} and the formalism developed by Straub et al.\cite{kim2010generalized,lu2013order} for the modeling of conventional phase transitions. We then use this knowledge to design novel alchemical perturbation potentials\cite{McCammon:Straatsma:92} and soft-core functions\cite{pitera2002comparison} which eliminate or reduce the adverse effects of the transitions. This advance leads to faster equilibration and to more robust binding free energy estimates for challenging systems hard to treat with conventional protocols.

In this context, an order/disorder transition occurs at a critical value of the alchemical progress parameter at which two conformational states are in equilibrium even though one of them has a much weaker ligand-receptor interaction energy than the other. The equilibrium is established because the high energy state is entropically favored relative to the low energy state to the same extent that it is disfavored energetically--again, in analogy with, for example, a gas being in equilibrium with its liquid even though the molar enthalpy of the gas is much less favorable than that of the liquid. Similarly, during binding free energy alchemical calculations, weakly coupled states are created in which ligand and receptor have much more conformational freedom and conformational entropy than more ordered coupled states. When it is not coupled to the receptor, the ligand can freely rotate and translate within the binding site region. Similarly, protein sidechains can experience a wide range of conformations when they are not forming interactions with the ligand. To transition from the unbound state to the bound state, the complex has to go through a tight entropic bottleneck related to the small likelihood to find a bound pose in the absence of ligand-receptor interactions; a frustrated process not unlike those of protein folding\cite{AlexeiV.Finkelstein2002,Zheng2007} and crystallization.\cite{yu2014order}

The equilibrium between ordered and disordered states is manifested in bimodal binding energy distributions with maxima separated by wide and poorly sampled binding energy gaps.\cite{Lapelosa2011} The binding free energy estimate is heavily influenced by the relative weight of competing modes.\cite{kilburg2018analytical} However, relative populations of ordered and disordered states can not be easily established because of the rare crossing events between them. The novel protocols we develop in this work are designed to reduce the binding energy gap and increase the number of binding/unbinding crossing events.

The paper is organized as follows. We first review the alchemical methodology and the analytic theory we employ. We then discuss methods to monitor and characterize the order/disorder transitions and ways to tune soft-core functions and alchemical perturbation functions to avoid them. We then illustrate the methods on two molecular complexes which display order/disorder biphasic behavior. The first is the complex between the L99A mutant of T4 lysozyme and 3-iodotoluene, a well known benchmarking system.\cite{gill2018binding} The second is the more challenging complex between the farnesoid X receptor and an inhibitor from a recent D3R grand challenge,\cite{gaieb2018d3r} which is more representative of typical medicinal applications. We show that in each case, order/disorder can be addressed leading to more efficient calculations and more robust binding free energy estimates.

\section{\label{sec:methods}Theory and Methods}

\subsection{Alchemical Transformations for Binding Free Energy Estimation}

We adopt a well-known statistical mechanics formulation of biomolecular non-covalent binding in which the standard free energy of binding is written as:\cite{Gilson:Given:Bush:McCammon:97,Boresch:Karplus:2003,Gallicchio2011adv}
\begin{equation}
\Delta G_b^\circ = \Delta G^\circ_\text{site} + \Delta G_b \label{eq:standard-free-energy}
\end{equation}
where
\begin{equation}
\Delta G^\circ_\text{site} = - \frac{1}{\beta} \ln C^\circ V_\text{site} \label{eq:Gzero-site}
\end{equation}
is the reversible work for transferring a ligand molecule from an ideal solution at the standard concentration
$C^\circ = 1 $M of ligand molecules, to a binding site region of volume $V_\text{site}$, and $\Delta G_b$ is the excess component corresponding to the desolvation process and the establishing of ligand-receptor interactions. In this work, the excess component is computed by means of an alchemical free energy perturbation schedule based on a potential energy function $U_\lambda (x)$, parametric on the alchemical progress parameter $\lambda$, which interpolates between the unbound state of the complex, described by the potential function $U_0(x)$, and that corresponding to the bound state, $U_1(x)$. When the solvent is modeled explicitly, a double-decouplng process\cite{Gilson:Given:Bush:McCammon:97,Chodera:Mobley:cosb2011} is used involving two alchemical legs, the first in which the ligand is decoupled from the solvent to reach a ``vacuum'' intermediate state, and a second leg in which the ligand is placed into the binding site and then coupled to the receptor.

Here we use a solvent potential of mean force formulation which allows to transfer the ligand directly from solution to the receptor binding site in a single alchemical process.\cite{Gallicchio2010,Gallicchio2011adv,kilburg2018assessment} In this formulation, in which the degrees of freedom of the solvent are averaged out, the effective potential energy function includes a standard molecular mechanics component describing covalent and non-bonded interactions plus an implicit solvation component to model hydration effects.\cite{Gallicchio:Levy:2004} In our implementation, the alchemical potential energy is expressed as
\begin{equation}
U_{\lambda}(x)=U_{0}(x)+W_{\lambda}(u)\label{eq:pert_pot}
\end{equation}
where $x$ represents the set of atomic coordinates of the molecular complex,
\begin{equation}
u(x)=U_{1}(x)-U_{0}(x) \label{eq:binding-energy}
\end{equation}
is the binding energy function, $U_{0}(x)$ is the effective potential energy of the uncoupled state, $U_{1}(x)$ is the effective potential energy of the coupled state, and $W_{\lambda}(u)$ is the alchemical perturbation function, which varies parametrically with $\lambda$. As defined in Eq.~(\ref{eq:binding-energy}), the binding energy function is defined as the potential energy difference between the coupled and uncoupled states of the complex in conformation $x$.\cite{Gallicchio2011adv}

In order to reproduce the physical coupled and uncoupled states of the complex at the beginning and end states of the alchemical transformation, it is necessary that the alchemical perturbation function is defined such that $W_{0}(u)=0$ and $W_{1}(u)=u$ at $\lambda=0$ and $\lambda=1$, respectively.  The linear function $W_{\lambda}(u)=\lambda u$ satisfies this requirement and is the standard choice for the alchemical perturbation function.\cite{Su:Gallicchio:Levy:2007} To obtain the binding free energy, set of samples of the binding energies, $u_i$, are collected during molecular dynamics simulations performed at a sequence of $\lambda$ values between $0$ and $1$. The excess free energy profile as a function of $\lambda$, $\Delta G_b(\lambda)$, is obtained by multi-state reweighting\cite{Shirts2008a} using the UWHAM method.\cite{Tan2012} The excess free energy of binding in Eq.~(\ref{eq:standard-free-energy}) is by definition the value of free energy profile at $\lambda=1$, $\Delta G_b = \Delta G_b(1)$.

The binding-energy representation of the perturbation energy used in this work has the advantage that an analytic theory is available to describe the statistics of the binding energy function and, by variable transformation, of any other quantity, such as the alchemical perturbation energy, which depends on it (see below).\cite{kilburg2018analytical} 
In the following, we exploit the theory to analyze order/disorder transitions along the alchemical path and derive soft-core functions and alchemical perturbation functions with superior replica exchange efficiency the conventional linear perturbation function $W_\lambda(u) = \lambda u$.

\subsection{Analytic Theory of Alchemical Molecular Binding}

The theory of alchemical binding we recently developed\cite{kilburg2018analytical} provides analytic expressions for the probability densities of the binding energy and of the binding free energy profile as a function of the alchemical progress parameter $\lambda$ for any binding energy-based alchemical perturbation potential $W_\lambda(u)$. The parameters of the model are obtained by fitting the prediction of the model to the binding energy probability distributions extracted from atomistic simulations of the complex.

Briefly (see reference \onlinecite{kilburg2018analytical} for the full derivation), the central quantity of the model is $p_0(u)$, the probability density of the binding energy $u$ in the uncoupled ensemble ($\lambda=0$ in this context).\cite{Gallicchio2010,Gallicchio2011adv} The model is based on the assumption that the statistics of the random variable $u$ is the superposition of two processes, one that describes the sum of many ``soft'' background ligand-receptor interactions and that follows central limit statistics, and another process that describes ``hard'' atomic collisions and that follows max statistics. 
The probability density $p_0(u)$ is expressed as the superposition of probability densities of a small number of binding modes  
\begin{equation}
p_0(u) = \sum_i c_i p_{0i}(u) \label{eq:superp}
\end{equation}
where $c_i$ are adjustable weights summing to $1$ and $p_{0i}(u)$ is the probability density corresponding to mode $i$ described analytically as (see reference \onlinecite{kilburg2018analytical} and appendix \ref{sec:a-pc} for the derivation):
\begin{eqnarray}
&&  p_{0i}(u) = p_{bi} g(u;\bar{u}_{bi},\sigma_{bi}) \nonumber \\
&& + (1-p_{bi}) \int_{0}^{+\infty}p_{WCA}(u';n_{li},\epsilon_{i},\tilde{u}_i)g(u-u';\bar{u}_{bi},\sigma_{bi})du' \label{eq:p0(u)conv2}
\end{eqnarray}
where $g(u;\bar{u},\sigma)$ is the normalized Gaussian density function of mean $\bar{u}$ and standard deviation $\sigma$ and
\begin{equation}
  p_{WCA}(u;n_l , \epsilon, \tilde{u}) = n_{l}\left[1-\frac{(1+x_{C})^{1/2}}{(1+x)^{1/2}}\right]^{n_{l}-1} 
  \frac{H(u)}{4\epsilon_{LJ}}\frac{(1+x_{C})^{1/2}}{x(1+x)^{3/2}}\, , \label{eq:pwcaf}
\end{equation}
where $x = \sqrt{u/\epsilon+\tilde{u}/\epsilon+1}$ and $x_C = \sqrt{\tilde{u}/\epsilon+1}$. The model for each mode $i$ depends on a number of adjustable parameters corresponding to the following physical quantities\cite{kilburg2018analytical}:
\begin{itemize}
\item $c_i$: relative population of binding mode $i$
\item $p_{bi}$: probability that no atomic clashes occur while in binding mode $i$
\item $\bar{u}_{bi}$: the average background interaction energy of binding mode $i$
\item $\sigma_{bi}$: the standard deviation of background interaction energy of binding mode $i$
\item $n_{li}$: the effective number of statistical uncorrelated atoms of the ligand in  binding mode $i$
\item $\epsilon_{i}$: the effective $\epsilon$ parameter of an hypothetical Lennard-Jones interaction energy potential describing the receptor-ligand interaction in binding mode $i$
\item $\tilde{u}_i$: the binding energy value above which the collisional energy is not zero in binding mode $i$
\end{itemize}
The parameters above, together with the weights $c_i$, are varied to fit the binding energy distributions obtained from numerical simulations\cite{kilburg2018analytical} (see Fig.~\ref{fig:plambda-predictions} and Table \ref{tab:parameters} for examples).  

All other quantities of the alchemical transformation can be obtained from $p_0(u)$.\cite{Gallicchio2010,Gallicchio2011adv} In particular, given $p_0(u)$, the probability density for the binding energy $u$ for the state with perturbation potential $W_\lambda (u)$ is 
\begin{equation}
p_{\lambda}(u)=\frac{1}{K(\lambda)}p_{0}(u) \exp\left[-\beta W_\lambda (u)\right]\label{eq:plambdau_1}
\end{equation}
where $\beta = 1/k_B T$,
\begin{equation}
K(\lambda)=\int_{-\infty}^{+\infty}p_{0}(u) \exp\left[-\beta W_\lambda (u)\right]  =\langle \exp\left[-\beta W_\lambda (u)\right] \rangle_{\lambda=0}\label{eq:plambdau_2}
\end{equation}
is the excess component of the equilibrium constant for binding and
\begin{equation}
\Delta G_b(\lambda) = - \frac{1}{\beta} \ln K(\lambda) \label{eq:gblambda}
\end{equation}
is the corresponding excess binding free energy profile. Note that for
a linear perturbation potential, $W_\lambda(u) = \lambda u$,
Eqs.~(\ref{eq:plambdau_2}) and (\ref{eq:gblambda}) state that the
binding free energy profile is related to the double-sided Laplace
transform of $p_\lambda(u)$.

\subsection{Order/Disorder Transitions along Alchemical Transformation}

In this context, an order/disorder phase transition occurs whenever two conformational states with significantly different average energies and entropies are in equilibrium. In these circumstances, transfer of population from one state to another takes place in a narrow range of the controlling thermodynamic parameter, typically temperature. For example, for the two-level system whereby one state is energetically favored, and the other is entropically favored, the width of the temperature transition $\delta T$ becomes increasingly narrow as the energy gap $\Delta E$ increases ($\delta T = 4 k_B T^2_m/\Delta E$ as measured by the derivative of the population of the upper level with respect to temperature at the midpoint temperature $T_m$).\cite{zuckerman2010statistical} In general, the hallmark of an order/disorder transition is the presence of a bimodal energy distribution with a sudden transfer of population from the low energy phase to the higher energy phase as the temperature is increased. The nature of temperature-activated order/disorder transitions and the conformational sampling bottlenecks they cause has been thoroughly investigated by John Straub and collaborators in a series of publications.\cite{Kim2010b,kim2010generalized,lu2013order,lu2014investigating}

In this work, we extend the analysis conducted by Straub et al.\ to the case of order/disorder transitions occurring along the alchemical path for binding.  Because there are many more ways that ligand atoms clash with receptor atoms than configurations free of clashes, at small $\lambda$ weakly coupled configurations are entropically favored relative to coupled states. Conversely, fully coupled configurations, which are free of severe clashes and are stabilized by favorable intermolecular interactions, predominate at large values of $\lambda$. Analogously to the temperature-driven transitions, coupled and weakly coupled states separated by a sizeable energy gap, are in equilibrium at a critical value of the alchemical parameter $\lambda$. The occurrence of an order/disorder equilibrium is evident from the presence of distinct modes of the binding energy distribution separated by a large binding energy gap. Hence, as in the case of variations in temperature, we identify order/disorder transitions by looking for $\lambda$-states with binding energy probability densities $p_\lambda (u)$ with two or more modes.

By setting to zero the derivative of the logarithm of Eq.~(\ref{eq:plambdau_1}), we find that the stationary points of $p_\lambda (u)$ obey the relationship
\begin{equation}
\lambda_{0}(u) - \frac{\partial W_{\lambda}(u)}{\partial u} = 0 \label{eq:lambda-function-condition}
\end{equation}
where we have introduced the $\lambda$-function
\begin{equation}
\lambda_{0}(u) \equiv \frac{1}{\beta}\frac{\partial\ln p_{0}(u)}{\partial u}\label{eq:lambda-function}
\end{equation}
of the system. As defined, the $\lambda$-function is analogous to the energy-dependent statistical temperature function $T_S(E)$ in the analysis of Straub et al.\cite{Kim2010b} for the canonical ensemble. Like the density of states $\Omega(E)$ and the statistical temperature $T_S(E)$ in the canonical ensemble, and $p_{0}(u)$ in the alchemical ensemble considered here, $\lambda_{0}(u)$ is a fundamental property of the molecular system and is independent of $\lambda$ and of the specific alchemical perturbation. Since in this model $p_0(u)$ is expressed analytically, the $\lambda$-function is also obtained in analytical form using Eq.~(\ref{eq:lambda-function}).

\begin{figure*}
\includegraphics[scale = 0.75]{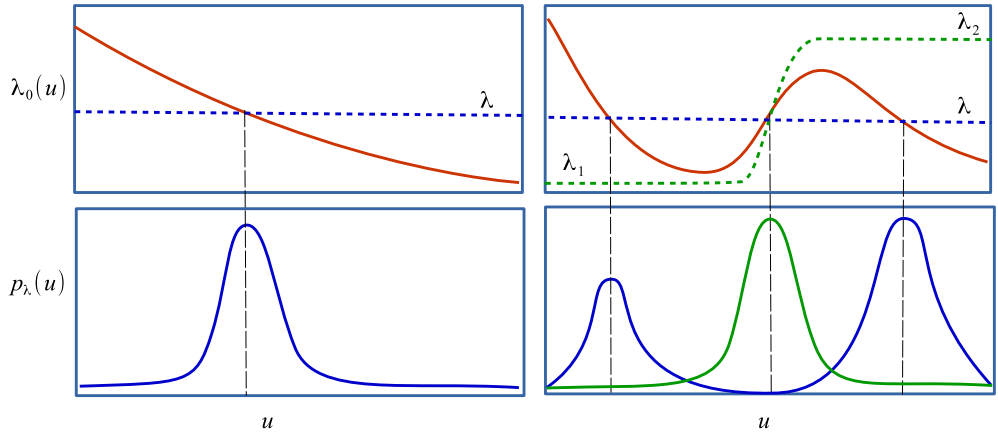}
\caption{Illustration of the relationship between the $\lambda$-function $\lambda_{0}(u)$ and the stationary points of the distributions of binding energies as a function of the alchemical parameter $\lambda$. The stationary points occur where $\lambda_{0}(u)$ (orange lines) intersects the function $\partial W_{\lambda}(u)/\partial u$. In the case of a linear perturbation, $W_{\lambda}(u)=\lambda u$, the latter is represented by an horizontal line at $\lambda$ (blue dashed lines).  When $\lambda_{0}(u)$ varies monotonically (left panel), it intersects the horizontal line at one point corresponding to the maximum of $p_{\lambda}(u)$ (blue, lower panels). Near an order/disorder transition (right panels), $\lambda_{0}(u)$ undergoes back-bending and intersects the $u=\lambda$ line (blue dashed line) at three points corresponding to two maxima and one in-between minimum of $p_{\lambda}(u)$ (blue). The bimodal behavior of $p_{\lambda}(u)$ can be converted a single mode (green) by working with a non-linear perturbation function $W_{\lambda}(u)$ whose derivative (green dashed line) intersects $\lambda_{0}(u)$ at only one point. \label{fig:phase-transition}}
\end{figure*}

Eq.~(\ref{eq:lambda-function-condition}) leads to the graphical construction illustrated in Fig.~\ref{fig:phase-transition} to locate the maxima and minima of $p_\lambda (u)$.\cite{Kim2010b} In the case of a linear perturbation ($W_{\lambda}(u)=\lambda u$) the stationary points of the distribution of binding energies $u$ at $\lambda$ occur when $\lambda_{0}(u)=\lambda$, that is when the $\lambda$-function intersects a horizontal line corresponding to the set value of $\lambda$ (Figure \ref{fig:phase-transition}).  For a general perturbation function $W_{\lambda}(u)$, the stationary points occur when $\lambda_{0}(u)$ intersects the function $\partial W_{\lambda}(u)/\partial u$ as in Eq.~(\ref{eq:lambda-function-condition}). As illustrated in Figure \ref{fig:phase-transition}, a order/disorder transition occurring with the linear perturbation function can be avoided by using a suitable, non-linear, perturbation function.\cite{Kim2010b}

The graphical construction in Figure \ref{fig:phase-transition} also easily indicates the regions of binding energies where $p_\lambda(u)$ in increasing or decreasing. When $\lambda_{0}(u) > \partial W_{\lambda}(u)/\partial u > 0$, that is when the curve representing the derivative of the perturbation function is below the $\lambda$-function, $p_\lambda(u)$ has a positive derivative and increases with increasing $u$. Conversely, when $\partial W_{\lambda}(u)/\partial u$ is above the $\lambda$-function, $p_\lambda(u)$ is a decreasing. These behaviors are confirmed by our results as shown in Figure \ref{fig:lambdaf-t4l}.

\subsection{Soft-Core Binding Energy Functions to Reduce Order/Disorder Transitions}

The alchemical potential of Eq.~(\ref{eq:pert_pot}) with the energy function $u(x)$ defined by Eq.~(\ref{eq:binding-energy}) leads to an unstable free energy estimation near the decoupled state.\cite{Steinbrecher2011} This issue, which is generally known as the ``end-point catastrophe'',\cite{pitera2002comparison} is due to the singularity of the derivative of the free energy profile near the decoupled state.\cite{simonson1993free} For a linear bias, $W_\lambda(u) = \lambda u$, it is straightforward to show from Eq.~(\ref{eq:pert_pot}) that the derivative of the binding free energy profile is the average binding energy
\begin{equation}
\frac{d\Delta G_b(\lambda)}{d \lambda} = \langle u \rangle_\lambda \, .
\end{equation}
Near the decoupled state, when ligand and receptor atoms interact weakly, it is very likely to encounter configurations in which ligand atoms and receptor atoms clash, leading to very unfavorable binding energies. The average binding energy and first derivative of the binding free energy profile diverge near the decoupled state at $\lambda = 0$. (This behavior can also be confirmed by considering the first moment of $p_0(u)$ using Eqs.~(\ref{eq:superp}) and (\ref{eq:p0(u)conv2}).)

To address the end-point singularity, we introduce a soft-core binding energy functions that caps the binding energy to some maximum value $u_{{\rm max}}$:\cite{Buelens2012}
\begin{equation}
u_{{\rm sc}}(u)=\begin{cases}
u & u\le 0\\
u_{{\rm max}} f_\text{sc}(y) & u>0
\end{cases}\label{eq:soft-core-general}
\end{equation}
where $u$ is the binding energy function, $y = u/u_{{\rm max}}$, and $f_\text{sc}(y)$ is a function that smoothly goes from zero at $y=0$ to one as $y$ goes to infinity (see below). The soft-core binding energy function $u_{{\rm sc}}=u_{{\rm sc}}(u)$ can be interpreted as a map from the unbound domain $[-\infty,+\infty]$ of $u$ to the $[-\infty,u_{{\rm max}}]$ domain of $u_{{\rm sc}}$, which is bounded from above at $u_{\max}$. The end states and their free energy difference are virtually unaffected by the replacement of $u$ with $u_{{\rm sc}}$. At the coupled state the binding energy distribution lays almost exclusively at negative binding energy values where $u$ and $u_{{\rm sc}}$ are the same [see Eq.~(\ref{eq:soft-core-general})]. At the uncoupled state ($\lambda = 0$) on the other hand, the biasing potential is zero regardless of the form of the binding energy function.

We consider two choices for $f_\text{sc}(y)$: a hyperbolic tangent
\begin{equation}
f_\text{sc}(y) = \tanh(y) \label{eq:tanh-sc}
\end{equation}
and a rational function designed to lessen the strength of order/disorder transitions
\begin{equation}
f_\text{sc}(y) = \frac{z^{a}-1}{z^{a}+1} \label{eq:rat-sc} \, ,
\end{equation}
where $z=1+2 y/a + 2 (y/a)^2$ and $a$ is an adjustable dimensionless exponent.  Both forms of $f_\text{sc}(y)$ above are invertible and lead to a $C(2)$-smooth soft-core binding energy function suitable for molecular dynamics applications when included in Eq.~(\ref{eq:soft-core-general}).

\begin{figure}
  \includegraphics{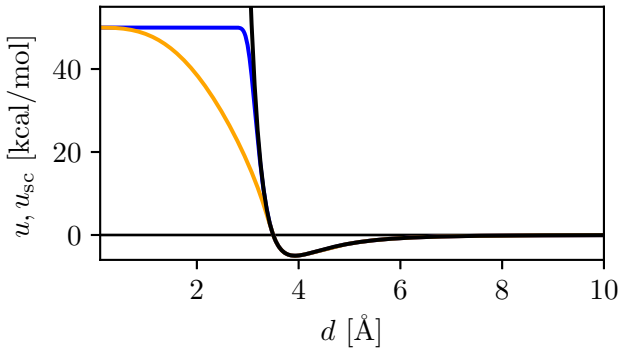}
  \caption{Illustration of the effect of the soft-core functions on the interaction energy between two atoms described by Lennard-Jones potential ($\epsilon_{{\rm LJ}}=5$ kcal/mol, $\sigma_{{\rm LJ}}=3.5$ \AA, black curve) as function of distance $r$. Hyperbolic tangent soft-core function, Eqs.~(\ref{eq:soft-core-general}) and (\ref{eq:tanh-sc}), with $u_{{\rm max}}=50$ kcal/mol (blue curve). Rational soft-core function, Eqs.~(\ref{eq:soft-core-general}) and (\ref{eq:rat-sc}), with $u_{{\rm max}}=50$ kcal/mol and $a=1/8$ (orange curve). While the LJ potential grows rapidly to infinity as $d^{-12}$ at short distances, the soft-core versions plateau at $u_{{\rm max}}$. The rational soft-core function provides a smoother transition and a smaller plateau region.
\label{fig:soft-core-LJ} }
\end{figure}

The soft-core binding energy function of Eqs.~(\ref{eq:soft-core-general})--(\ref{eq:rat-sc}) replaces the binding energy function $u$ in the binding energy-based alchemical potential of Eq.~(\ref{eq:pert_pot}). As an illustration, Fig.~(\ref{fig:soft-core-LJ}) shows the effect of the soft-core function on the binding energy between two particles interacting by a Lennard-Jones interaction. While both soft-core functions cap the binding energy function to $u_{\rm max}$, the rational soft-core function leads to a smaller high-energy plateau region which, as shown below, reduces the strength of order/disorder transitions near the decoupled states (see below).

\subsubsection{Probability Density and $\lambda$-Function of Soft-Core Binding Energy}

The soft-core binding energy function (\ref{eq:soft-core-general}) is interpreted as a redefinition of the binding energy of the system.  Application of the alchemical theory for the soft-core definition of the binding energy hinges on knowledge the probability density, $p_{0}^{{\rm sc}}(u_{{\rm sc}})$, of $u_{{\rm sc}}$ at $\lambda=0$.  This can be obtained from the probability density, $p_{0}(u)$, of $u$ using the variable transformation formula:
\begin{equation}
p_{0}^{{\rm sc}}(u_{{\rm sc}})=p_{0}[u(u_{{\rm sc}})]u'(u_{{\rm sc}})\label{eq:p0-sc}
\end{equation}
where $u=u(u_{{\rm sc}})$ is the inverse of the soft-core function [Eq.~(\ref{eq:soft-core-general})]. Since $p_0(u)$ is available in analytical form [Eq.~(\ref{eq:superp})] and Eqs.~(\ref{eq:tanh-sc}) and (\ref{eq:rat-sc}) are analytically invertible and differentiable, Eq.~(\ref{eq:p0-sc}) provides an analytical expression of the probability density of the soft-core binding energy at $\lambda=0$. 

To $\lambda$-function for the soft-core binding energy is obtained by differentiating the logarithm of Eq.~(\ref{eq:p0-sc}) with respect to $u_{sc}$, yielding
\begin{equation}
\lambda^{\rm sc}_{0}(u_{{\rm sc}})=\frac{1}{\beta}\frac{\partial\ln p_{0}^{{\rm sc}}(u_{{\rm sc}})}{\partial u_{{\rm sc}}}=\frac{1}{\beta}\left[\lambda_{0}(u)u'(u_{{\rm sc}})+\frac{u''(u_{{\rm sc}})}{u'(u_{{\rm sc}})}\right]
\label{eq:lambda-function-sc}
\end{equation}
where $\lambda_{0}(u)$ is the $\lambda$-function of the binding
energy without soft-core {[}Eq.~(\ref{eq:lambda-function}){]} evaluated
at $u=u(u_{{\rm sc}})$.

It can be shown from Eq.~(\ref{eq:lambda-function-sc}), and Eqs.~(\ref{eq:soft-core-general}), (\ref{eq:tanh-sc}), and (\ref{eq:rat-sc}), that the $\lambda$-functions with both of the soft-core functions we are considering diverges to $+\infty$ at $u_{\rm sc} = u_{\rm max}$ where the first derivative of the soft-core function is zero. Interestingly, the characteristics of the singularity are system-independent (the term $\lambda_{0}(u)u'(u_{{\rm sc}})$ is zero at the singularity) and they depend only on the nature of the soft-core function. As illustrated below, the divergence of $\lambda^{\rm sc}_{0}(u_{{\rm sc}})$, implies that all distribution functions of the soft-core binding energy, regardless of the form of the perturbation potential and of the alchemical schedule, will present a stationary point (a minimum, in fact) near $u_{\rm sc} = u_{\rm max}$. Because, $\partial w_\lambda(u)/\partial u$ is finite for any well-behaved perturbation function, near $p_\lambda(u)$ is necessarily an increasing function near $u_{\rm sc} = u_{\rm max}$.  However, in practice the upward trend of $p_\lambda(u)$ may not be observed in alchemical states, such as bound states, for which the population near $u_{\rm max}$ is negligibly small.

\subsection{Alchemical Perturbations Functions to Reduce Order/Disorder Transitions}

As shown below, while suitable soft-core functions can reduce the binding energy gap across an order/disorder transition, they can also create order/disorder sampling bottlenecks elsewhere along the alchemical path (see for example Figure \ref{fig:lambda-functions-t4l}). We found that combining a soft-core function with new kinds of perturbation potentials to be particularly beneficial. One of the most effective perturbation potentials that we identified is
\begin{equation}
  W_{\lambda}(u)=\frac{\lambda_{2}-\lambda_{1}}{\alpha}\ln\left[1+e^{-\alpha(u-u_{0})}\right]+\lambda_{2}u+w_{0}
  \label{eq:ilog-function}
\end{equation}
which we named the integrated logistic biasing function. The parameters $\lambda_{2}$, $\lambda_{1}$, $\alpha$, $u_{0}$, and $w_{0}$ are functions of $\lambda$. The name of this function comes from the fact that its derivative is the logistic function (also know as Fermi's function)
\begin{equation}
  \frac{\partial W_{\lambda}(u)}{\partial u}=\frac{\lambda_{2}-\lambda_{1}}{1+e^{-\alpha(u-u_{0})}}+\lambda_{1}
  \label{eq:ilog-function-der}
\end{equation}
which is sketched out in Figures \ref{fig:phase-transition} and \ref{fig:lambdaf-t4l}(A).

The parameter $\lambda_{1}$ is the height of the horizontal branch at the low binding energy end, and $\lambda_{2}$ is the height at the high binding energy end. The parameter $u_{0}$ controls the position of the switch from $\lambda_{1}$ to $\lambda_{2}$, and $\alpha$ controls the range of the switch. The parameter $w_{0}$ is an overall energy offset. As illustrated below, bimodal behavior can be avoided by properly tuning the parameters of the integrated logistic function in regions of the alchemical path affected by order/disorder transitions.  Conversely, the integrated logistic biasing function behaves as a linear biasing function away from the transition region.

\subsection{Hamiltonian Replica Exchange Conformational Sampling}

In this work, we employ the Hamiltonian Replica Exchange algorithm\cite{Sugita2000,Felts:Harano:Gallicchio:Levy:2004,Ravindranathan:Gallicchio:Levy:2006,Okumura2010} in alchemical space to accelerate conformational sampling.\cite{Woods2003,Rick2006,Gallicchio2010} In the context of the alchemical potential energy function (\ref{eq:pert_pot}), Hamiltonian Replica Exchange consists of performing MD of multiple replicas of the system each assigned a value of the $\lambda$ parameter. Periodically, the assignment of replicas to $\lambda$ states is varied in such a way so as to preserve a canonical distribution of conformations at each $\lambda$. The algorithm allows each replica to explore a wide range of $\lambda$-states, from coupled to uncoupled, thereby accelerating the sampling of receptor-ligand intermolecular degrees of freedom.  For the calculations reported here, we have employed the asynchronous implementation of Replica Exchange (ASyncRE)\cite{gallicchio2015asynchronous} with the Gibbs Independence Sampling algorithm\cite{chodera2011replica} for state reassignments (see Appendix \ref{sec:gibbs}).

The occurrence of an order/disorder phase transition has been shown to limit the rate at which replicas diffuse in thermodynamic space,\cite{Kim2010b} and hinder the ability of replica exchange to accelerate conformational sampling. This is because, at a phase transition, nearby thermodynamic states can be separated by a large energy gap. In the alchemical case, the probability that two replicas exchange their $\lambda$-states is large when the two replicas have similar binding energies, and it decreases rapidly when they are separated by a large binding energy gap. This effect is best appreciated in the case of a linear perturbation potential $W_\lambda (u) = \lambda u$ for two replicas on opposite sides of the order/disorder transition. The replica in the ordered state with more favorable binding energy is preferentially at the larger value of $\lambda$ whereas the replica in the disordered state is more likely to have higher binding energy at a smaller value of $\lambda$. In this case the probability of exchange, which is proportional to the factor $\exp(\Delta \lambda \Delta u)$, where $\Delta u > 0$ is the difference in binding energies and $\Delta \lambda < 0$ the difference in $\lambda$ values between the two replicas, decreases exponentially with increasing binding energy separation. In this work we show that Hamiltonian RE efficiency can be improved by using an alchemical perturbation potential, such as the one proposed above, that removes or softens order/disorder transitions.

\subsection{Computational Details}

\begin{figure*}
  \centering
  (A)\begin{minipage}[t]{0.45\textwidth}
    \ \newline
    \includegraphics[scale=0.17]{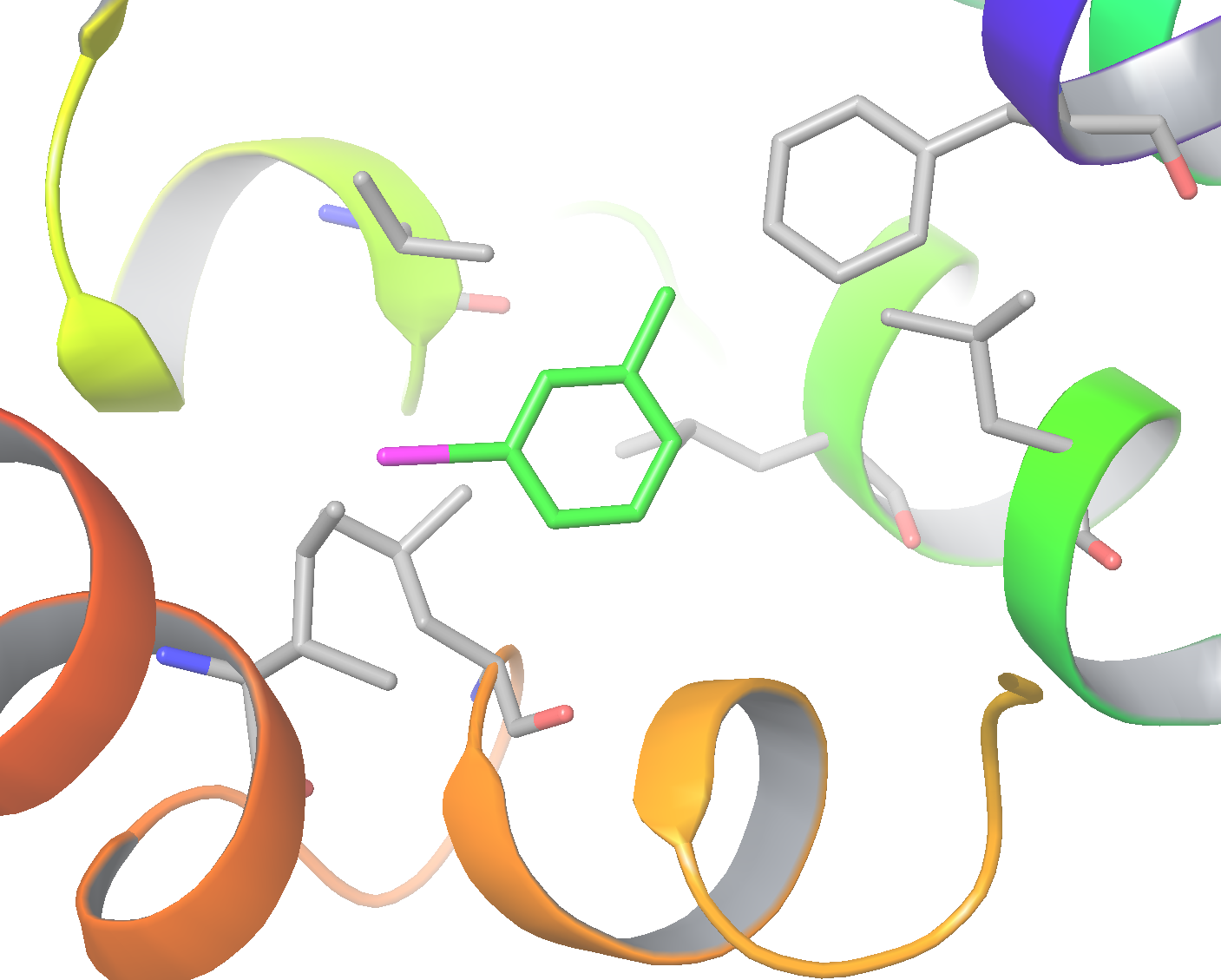}
  \end{minipage}
  (B)
  \begin{minipage}[t]{0.45\textwidth}
    \ \newline
    \includegraphics[scale=0.15]{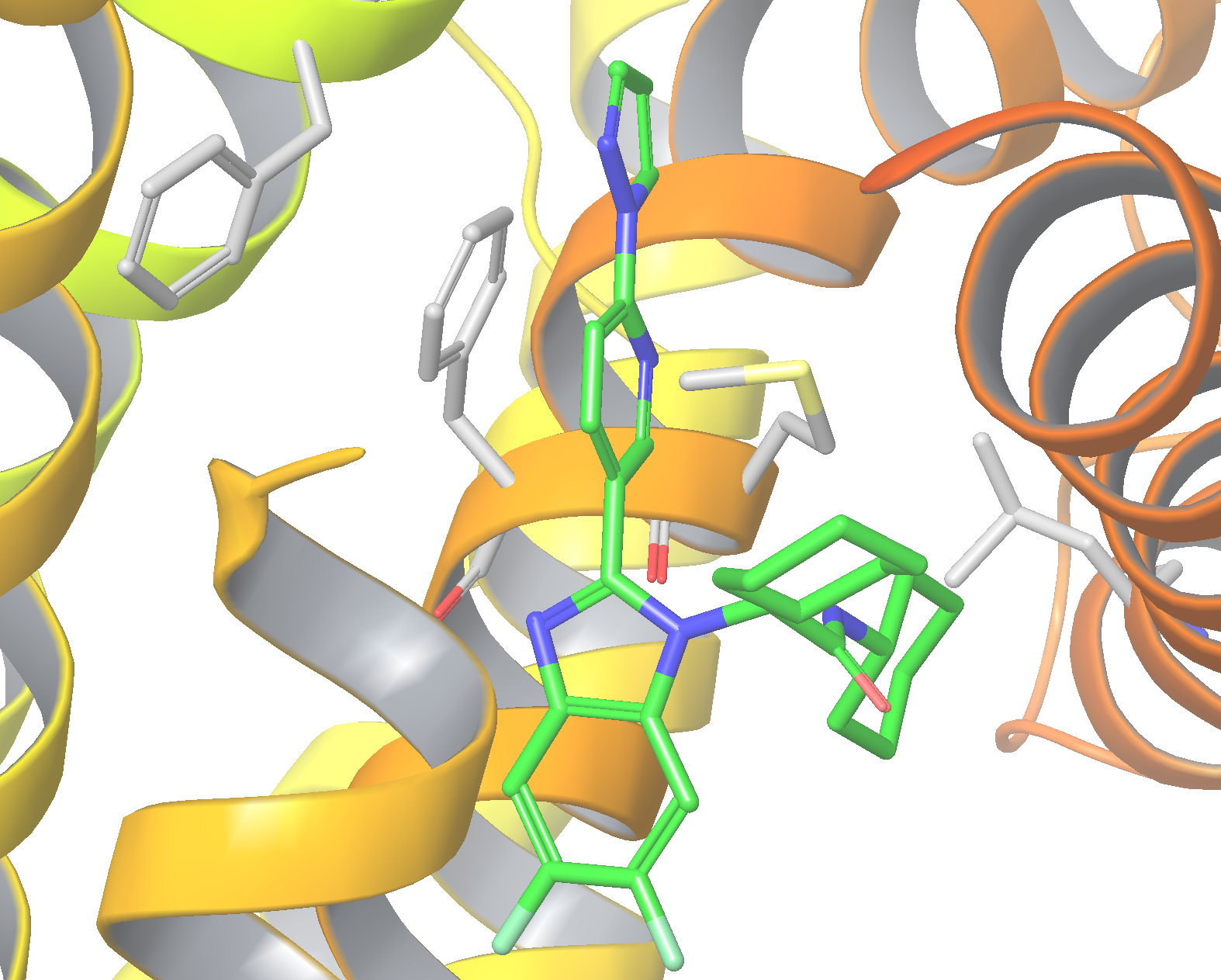}
  \end{minipage}
  \caption{(A) The complex between the L99A mutant of T4 lysozyme (T4L) and 3-iodotolune and (B) the complex between the farnesoid X receptor (FXR) and ligand FXR-26. In each structure, the carbon atoms of the ligand are colored green and some of the residues lining the binding pockets are shown.
    \label{fig:structures}
  }
\end{figure*}

The molecular structures of the L99A T4 lysozyme receptor (PDB ID 4W53) and of the Farnesoid X Receptor (FXR), see Figure \ref{fig:structures}, were prepared from their crystal structures\cite{gaieb2018d3r,merski2015homologous} using the protein preparation wizard of Maestro (Schrodinger Inc) with default settings.  Residues 1 through 71 of the T4 lysozyme receptor, which do not participate in ligand binding, were removed. The positions of the C$\alpha$ atoms of the receptors were loosely restrained using a flat-bottom harmonic potential with a tolerance of $1.5$ \AA. 3-iodotoluene was placed in the binding site of T4L by superimposing it on the structure of bound toluene. The FXR-26 inhibitor bound to FXR was placed based on the available crystal structure.\cite{gaieb2018d3r} Both ligands were prepared using the LigPrep facility of Maestro at pH 7. The single-decoupling binding free energy simulations were set up using the Single Decoupling Method (SDM) workflow ({\tt github.com/egallicc/openmm\_sdm\_workflow}). OPLS-AA 2005 force field parameters\cite{Kaminski:2001,Banks:Gallicchio:Levy:2005} were assigned using Desmond. The binding site volume for the T4L system is defined as any conformation in which the ligand center of mass is within $2.5$ {\AA} of the center of mass of the $\mathrm{C_{\alpha}}$ atoms of residues 79, 84, 88, 91, 96, 104, 112, 113, 122, 133, and 150 of the T4L receptor. The binding site volume for the FXR system is similarly defined using the center of mass of the $\mathrm{C_{\alpha}}$ atoms of residues 273, 277, 291, 333, 336, 340, 356, and 369 of the FXR receptor.  In both systems, the ligand was sequestered within the binding site by means of a flat-bottom harmonic potential with a force constant of $25$ kcal/mol {\AA}$^2$ applied to atoms with distances greater than $2.5$ {\AA}. With these settings, the value of $\Delta G^\circ_{\rm site}$ in Eq.~(\ref{eq:Gzero-site}) is $1.97$ kcal/mol.

SDM alchemical calculations employed the OpenMM\cite{eastman2017openmm} MD engine with the AGBNP ({\tt github.com/\-egallicc/\-openmm\_agbnp\_plugin}) and SDM integrator plugins ({\tt github.com/\-rajatkrpal/\-openmm\_sdm\_plugin.git}) using the OpenCL platform. The ASyncRE software,\cite{gallicchio2015asynchronous} customized for OpenMM and SDM ({\tt github.com/\-baofzhang/\-async\_re-openmm.git}), was used for the Hamiltonian Replica Exchange in $\lambda$ space. We used 16 and 24 replicas, respectively, for the T4L and FXR complexes, set at equally spaced values of $\lambda$ between $0$ and $1$. The $\lambda$-dependent parameters, listed in Tables \ref{tab:ilog-t4l}, \ref{tab:ilog-fxr-umax50}, and \ref{tab:ilog-fxr-umax100}, of the integrated logistic schedule were chosen so as to avoid the order/disorder transitions near $u=0$, as described in the Results section. Molecular dynamics runs were conducted for a minimum of 6 ns per replica with a 1 fs time-step at $300$ K, exchanging approximately every 10 to 20 ps. A Langevin thermostat at $300$ K with a relaxation time constant of $2$ ps was used. Binding energy samples and trajectory frames were recorded every 5 ps. The calculations were performed on a farm of GPU servers at Brooklyn College and on the XSEDE Comet GPU HPC cluster using a mix of 780 Ti, 2080, Titan Xp, K80, and P100 NVIDIA GPUs.

\section{\label{sec:results} Results}


The adverse effects of order/disorder transitions in alchemical calculations, and the ability of suitably crafted soft-core functions and perturbation potentials to avoid or ameliorate them, are illustrated here for two systems: the complex of L99A T4 lysozyme with 3-iodotoluene (T4L99A/3-iodotoluene) and the complex between the farnesoid X receptor with inhibitor 26 (FXR/26)\cite{gaieb2018d3r} (Figure \ref{fig:structures}). The T4 lysozyme receptor is often used to test alchemical and conformational sampling methods.\cite{Boyce2009,Gallicchio2010,gill2018binding} The FXR system is a much more challenging system and more representative of those encountered in medicinal applications. Both of these systems display order/disorder transitions along the alchemical path. The T4L/3-iodotoluene complex was simulated with the hyperbolic tangent (TanhSC), and the rational function soft-core (RatSC) functions with a linear perturbation schedule as well as with the integrated logistic perturbation potential (ILog) with the RatSC soft-core function. The FXR/26 complex was simulated with the same settings as T4L99A/3-iodotoluene and with an additional soft-core parameterization. In total, we report the results of 3 alchemical calculations for T4L/3-iodotoluene and 6 alchemical calculations for FXR/26 (see Table \ref{tab:simulations} for a summary).

\begin{table*}
  \caption{\label{tab:simulations}Summary of the alchemical calculations reported as part of this work}
\begin{ruledtabular}
\begin{tabular}{lcccccc}
Simulation id       &   soft-core function  & $u_{\rm max}$& $a$    &  biasing potential   & number of replicas & simulation length\tabularnewline
\multicolumn{7}{c}{T4L99A/3-iodotoluene}\tabularnewline
1                   & TanhSC                & $50$       &  --    & Linear               & $16$               & $6$ ns           \tabularnewline
2                   & RatSC                 & $50$       & $1/16$ & Linear               & $16$               & $6$ ns           \tabularnewline
3                   & RatSC                 & $50$       & $1/16$ & ILog                 & $16$               & $6$ ns           \tabularnewline
\multicolumn{7}{c}{FXR/26}\tabularnewline
1                   & TanhSC                & $50$       &  --    & Linear               & $24$               & $6$ ns           \tabularnewline
2                   & RatSC                 & $50$       & $1/16$ & Linear               & $24$               & $6$ ns           \tabularnewline
3                   & RatSC                 & $50$       & $1/16$ & ILog                 & $24$               & $6$ ns           \tabularnewline
4                   & TanhSC                & $100$      &  --    & Linear               & $24$               & $6$ ns           \tabularnewline
5                   & RatSC                 & $100$      & $1/14$ & Linear               & $24$               & $6$ ns           \tabularnewline
6                   & RatSC                 & $100$      & $1/14$ & ILog                 & $24$               & $6$ ns           \tabularnewline
\end{tabular} 
\end{ruledtabular}
\end{table*}

The results are organized as follows. We first present the parameterization of the analytical model for the two systems (Figure \ref{fig:plambda-predictions} and Table \ref{tab:parameters}) and the corresponding analytical predictions of the effects of different choices of soft-core functions and alchemical perturbations potentials (Figures \ref{fig:lambda-functions-t4l} and \ref{fig:lambda-functions-fxr}). We then present the design of the alchemical schedule with the integrated logistic perturbation potential to address order/disorder transitions (Figure \ref{fig:ilogistic-design}). The results of the alchemical calculations are shown next, with a particular focus on the effect of the various settings on the conformational sampling and replica exchange efficiency (Figures \ref{fig:plambdas}, \ref{fig:binde_traj}, and Table \ref{tab:binding_unbinding}), and the convergence of the binding free energies (Table \ref{tab:free_energies} and Figure \ref{fig:equilibration}).

\subsection{Parameterization of the Analytical Model of Alchemical Binding}

\begin{figure*}
  \centering
  (A)\begin{minipage}[t]{0.45\textwidth}
    \begin{center}
    T4L
    
    \includegraphics{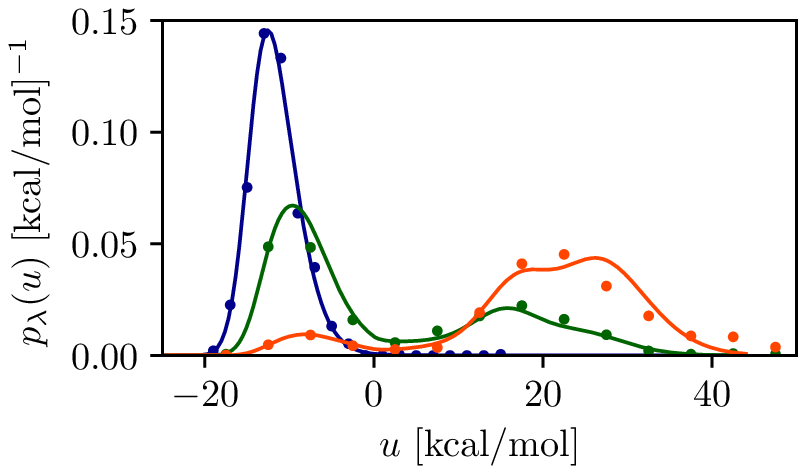}
    \end{center}
  \end{minipage}
  \hfill
  (B)
  \begin{minipage}[t]{0.45\textwidth}
    \begin{center}
      FXR
      
    \includegraphics{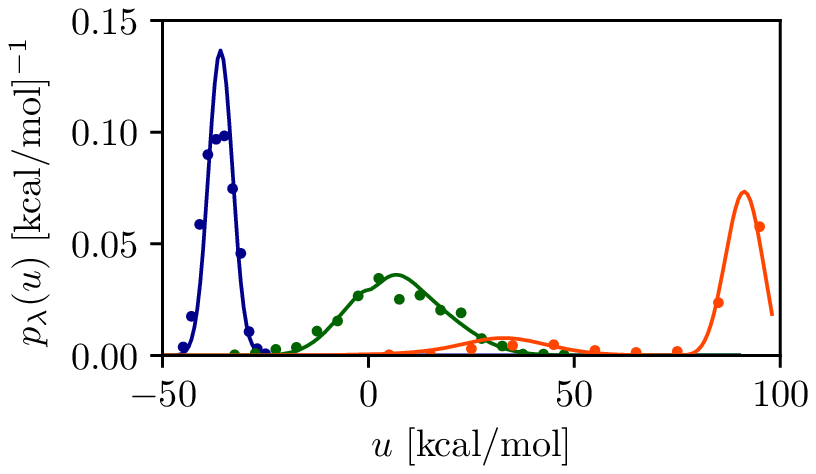}
    \end{center}
  \end{minipage}
\caption{\label{fig:plambda-predictions} The predicted (continuous lines) and observed (points) binding energy probability densities, $p_\lambda (u)$, for (A) the T4L/3-iodotoluene complex for the linear perturbation function $W_\lambda (u) = \lambda u$ and the rational soft-core function with $u_\text{max} = 50$ kcal/mol and $a = 1/16$ at $\lambda = 0.171$ (orange), $\lambda = 0.229$ (green), and $\lambda = 0.4$ (blue), and for (B) the FXR complex with the integrated logistic perturbation function [Eq.~(\ref{eq:ilog-function})] and the rational soft-core function with $u_\text{max} = 100$ kcal/mol and $a = 1/14$ at $\lambda = 0.174$ (orange), $\lambda = 0.304$ (green), and $\lambda = 0.696$ (blue). The predicted distributions are obtained using the analytical model for $p_0 (u_)$, Eqs.~(\ref{eq:superp}) and (\ref{eq:p0(u)conv2}), with the parameters listed in Table \ref{tab:parameters}}
\end{figure*}

The parameters for the analytical description of $p_0(u)$ [Eq.~(\ref{eq:superp})]\cite{kilburg2018analytical} for the complexes studied in this work are listed in Table \ref{tab:parameters}. The good level of agreement between the analytical binding energy distributions, from Eq.~(\ref{eq:plambdau_1}), and the histograms obtained from the alchemical calculations are illustrated in Figure \ref{fig:plambda-predictions}.

The model predicts the presence of three modes for the T4L/3-iodotoluene system and two modes for the FXR system (Table \ref{tab:parameters}). In the case of T4L/3-iodotoluene, two of the modes (modes 1 and 2) correspond to alternative binding poses of 3-iodotoluene, one with strong interactions with the receptor (mode 1, with $\bar{u}_{b} = -11$ kcal/mol) and another (mode 2), more weakly bound, but approximately 13 times more likely to occur than the first in the uncoupled ensemble at $\lambda = 0$ (see the corresponding values of the statistical weights in the second column of Table \ref{tab:parameters})). While they formally describe binding poses, the analytical model predicts that in the absence of receptor-ligand interactions the probability of occurrence of clash-free configurations is very small for these states (from $10^{-6}$ to $10^{-5}$, third column of Table \ref{tab:parameters}). The third mode of T4L/3-iodotoluene (mode 3, with $89$\% weight) corresponds to conformations in which the ligand is nearly freely rotating and translating within the binding site. We refer to modes such as this as ``unbound'', keeping in mind however that in the alchemical approach the ligand is not allowed to leave the binding site. As reflected by the negligible $p_b$ parameter, the highly unfavorable average binding energy parameter, and the larger $n_l$ parameter (columns 3, 4, and 8), in this mode clashes with receptor atoms are more severe and occur with an overwhelmingly large probability.

The analytical model predicts a single binding pose for the FXR/26 complex (mode 1 in Table \ref{tab:parameters}). The bound mode (mode 1) is favored relatively to the unbound mode (mode 2) by a more favorable interaction energy ($\bar{u}_{b} = -28$ vs.\ $10^5$ kcal/mol). At the same time, as measured by its statistical weight (second column in Table \ref{tab:parameters}), the bound mode is predicted to be one hundred thousand times less likely than the unbound mode (mode 2) at $\lambda = 0$. In addition, the probability of occurrence of configurations free of atomic clashes while in mode 1, is predicted to be very small ($1.5 \times 10^{-10}$, third column of Table \ref{tab:parameters}). Taking into account the small population of mode 1, the probability of observing conformations free of clashes at $\lambda=0$ is predicted to be as small as $1.5 \times 10^{-15}$. The logarithm of this number is a measure of the entropy loss for binding.

Inspection of the molecular dynamics trajectories largely confirms the results of statistical analysis above. Indeed, 3-iodotolune is observed to visit predominantly two binding poses related by a swap of the positions of the methyl and iodo substituents within the binding site. The pose with weaker interactions (mode 2) is entropically favored and occurs more often at smaller values of $\lambda$. Conversely, the pose with stronger interactions occurs more frequently at values of $\lambda$ closer to 1. Conversely, and in agreement with the predictions of the analytical model, the inhibitor is observed to visit only one binding pose. In all complexes examined and with any of the simulation settings, we observed that, while in one of the bound modes, the ligand oscillates around the stable binding pose and that the range of the oscillations is greater near the uncoupled state at $\lambda=0$. Infrequently, and only if $\lambda$ is smaller than a critical value, the ligand transitions to a disordered state where it explores a wide range of positions and orientations within the binding site. Replicas mirror this behavior in the disordered unbound mode which, infrequently and only if $\lambda$ is larger than a critical value, transition to one of the binding modes where the ligand is ordered. As shown below, the frequency of order/disorder transitions such as these is influenced by the choice of the soft-core function and of the alchemical perturbation schedule.

\begin{table*}
  \caption{\label{tab:parameters}Optimized parameters for the analytical model of binding for the three complexes studied in this work}
\begin{ruledtabular}
\begin{tabular}{lccccccc}
       & weight             & $p_b$            & $\bar{u}_{b}$\footnote{In kcal/mol} & $\sigma_{b}^{\rm a} $ & $\epsilon^{\rm a}$ & $\tilde{u}^{\rm a}$ & $n_l$\tabularnewline
\multicolumn{8}{c}{T4L99A/3-iodotoluene}\tabularnewline
mode 1 & $7.5 \times 10^{-3}$& $1\times 10^{-5}$ & $-11.0$                            & $1.95$             & $20$             &   $-20$            & $5.5$ \tabularnewline
mode 2 & $0.10$             & $1\times 10^{-6}$ & $ -4  $                            & $2.8$              & $20$             &   $-4$             & $5.5$ \tabularnewline
mode 3 & $0.8925$           & $0$              & $100$                              & $10$               & $100$            &   $100$            & $8$   \tabularnewline
\multicolumn{8}{c}{FXR/FXR-26}\tabularnewline
mode 1 & $1 \times 10^{-5}$  & $1.5 \times 10^{-10}$ & $-28$                        & $2.9$             & $6$             &   $-6$            & $19$ \tabularnewline
mode 2 & $\sim 1$$^b$           & $0$                  & $1 \times 10^{5}$             & $1 \times 10^{4}$   & $100$           &  $1 \times 10^{5}$  & $30$   \tabularnewline
\end{tabular} 
\end{ruledtabular}
\end{table*}

\subsection{$\lambda$-Functions}

\begin{figure}
  \includegraphics{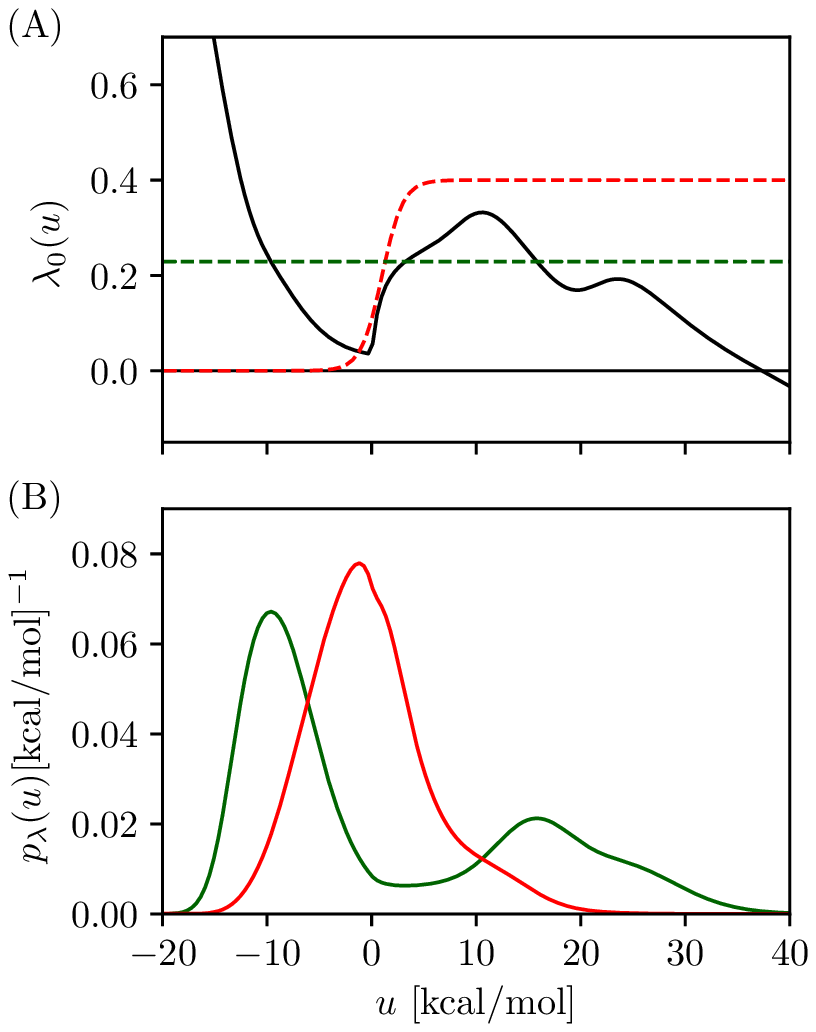}
  \caption{\label{fig:lambdaf-t4l} (A) The $\lambda$-function for the
    T4L/3-iodotoluene complex with the rational soft-core function
    with $u_{\rm max} = 50 $ kcal/mol and $a = 1/16$ (black) and
    $\partial W_\lambda(u)/\partial u$ for a linear perturbation
    function (green, dashed, $\lambda=0.229$) and for the integrated
    logistic function (red, dashed, $\lambda = 0.4$, $\lambda_1 = 0$,
    $\lambda_2 = 0.4$, $\alpha = 1$ $({\rm kcal/mol})^{-1}$, $u_0 = 1$
    kcal/mol, $w0 = -0.677$ kcal/mol. (B) The binding energy
    probability densities, $p_\lambda (u)$, predicted for the linear
    (green) and integrated logistic (red) perturbation functions with
    the parameters as in (A). The maxima and minima of the probability
    densities correspond to the intersections of the
    $\lambda$-function with the corresponding $\partial
    W_\lambda(u)/\partial u$ curves. The probability distribution with
    the linear perturbation is bimodal whereas the one with the
    integrated logistic potential has a single maximum. }
\end{figure}

\begin{figure}
  \includegraphics{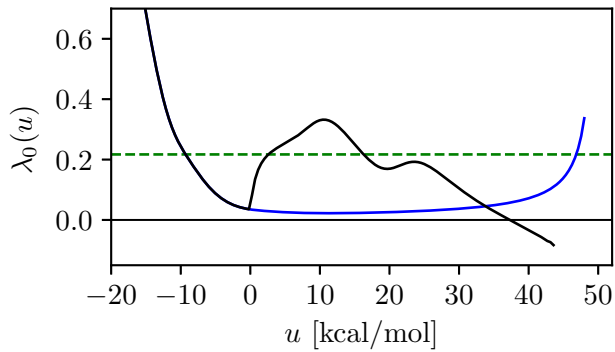}
  \caption{\label{fig:lambda-functions-t4l} The $\lambda$-function for the T4L/3-iodotoluene complex with the hyperbolic tangent soft-core function with $u_{\rm max}= 50$ kcal/mol (blue), and with the rational soft-core function with $u_{\rm max} = 50 $ kcal/mol and $a = 1/16$ (black). The two functions are equal for $u<0$ where the two soft-core functions are the same.  A linear $\lambda$-state (represented by the green dashed horizontal line) intersects the $\lambda$-functions where the corresponding maxima and minima of the binding energy distribution occur.}
\end{figure}

\begin{figure}
  \includegraphics{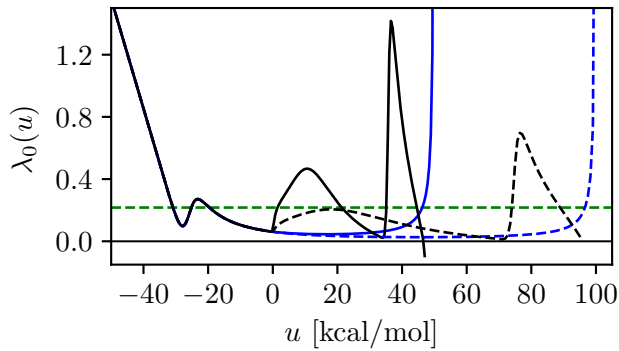}
  \caption{\label{fig:lambda-functions-fxr} The $\lambda$-functions
    for the FXR/26 complex with the hyperbolic tangent soft-core
    function with $u_{\rm max}= 50$ kcal/mol (blue, solid) and $u_{\rm
      max}= 50$ kcal/mol (blue, dashed), and with the rational
    soft-core function with $u_{\rm max} = 50 $ kcal/mol and $a =
    1/16$ (black, solid) and $u_{\rm max} = 100 $ kcal/mol and $a =
    1/14$ (black, dashed). The $\lambda$-functions are equal for $u<0$
    where the two soft-core functions are the same.  A linear
    $\lambda$-state (represented by the green dashed horizontal line)
    intersects the $\lambda$-functions where the corresponding maxima
    and minima of the binding energy distribution occur.}
\end{figure}

The $\lambda$ functions obtained from the analytical model [Eqs.~(\ref{eq:lambda-function}) and (\ref{eq:lambda-function-sc})] and the parameters in Table \ref{tab:parameters} are shown in Figures \ref{fig:lambdaf-t4l}, \ref{fig:lambda-functions-t4l}, and \ref{fig:lambda-functions-fxr}. As discussed above, the $\lambda$-function depends only on the chemical system and the choice of the soft-core function. Once parameterized, the analytical model yields, using Eq.~(\ref{eq:lambda-function-sc}), the $\lambda$ function for any choice of binding-energy based soft core function.

A non-monotonic behavior of the $\lambda$-function corresponds to multi-modal distributions which may be indicative of order/disorder transitions during the alchemical transformation. The data in Figure \ref{fig:lambda-functions-t4l} for the T4L/3-iodotoluene system shows that the linear schedule and the TanhSC soft-core function is expected to yield an order/disorder transition between an ordered state with negative average binding energy and a disordered state with average binding energy close to $u_{\rm max}$. These states correspond, for example, to the intersections in Figure \ref{fig:lambda-functions-t4l} of the horizontal line at $\lambda = 0.217$ with the $\lambda$-function (as discussed above, the intersection near the upper limit of $u$ corresponds to a minimum followed by a maximum at $u=u_{\rm max}$). In this particular case, we expect that the ordered state and the disordered state are separated by a large energy gap of more than $50$ kcal/mol which leads, as presented below, to rare transitions and poor replica exchange efficiency. With the RatSC soft-core function (black curve in Figure \ref{fig:lambda-functions-t4l}), in contrast, the energy gap between low and high binding energy states is significantly reduced. For instance, at $\lambda=0.217$ the high energy state of $p_\lambda(u)$ is predicted to peak at $u \simeq 17$ kcal/mol rather than at $50$ kcal/mol with the TanhSC soft-core. As a result, the binding energy gap of the order/disorder transition is predicted to be reduced by roughly half.

As shown in Figure \ref{fig:lambda-functions-fxr}, the FXR/26 system displays order/disorder transitions of similar nature but of greater strength and complexity than T4L/3-iodotoluene. As indicated by the presence with the RatSC soft-core function of an additional, strong, peak of the $\lambda$-function at high binding energies, we expect that the alchemical coupling process for FXR/26 to display multi-modal behavior. Specifically, we expect to observe, starting at $\lambda=0$, sharp transfers of population from a highly disordered state with binding energies close to $u_{\rm max}$ to a less disordered state with binding energies in the $10$--$40$ kcal/mol range and then to coupled states at negative binding energies. With the same RatSC soft-core function as T4L/3-iodotoluene capped at $u_{\rm max} = 50$ kcal/mol (solid black curve in Figure \ref{fig:lambda-functions-fxr}) we expect entropic bottlenecks against binding to occur at binding energies at around $0$ and $35$ kcal/mol, respectively. These bottlenecks can be identified graphically by locating the intersections between horizontal lines at $\lambda$ and the $\lambda$-function corresponding to minima of the binding energy distribution function. The strength of the order/disorder transitions can be reduced using a less aggressive soft-core function with $u_{\rm max} = 100$ kcal/mol (dashed black curve in Figure \ref{fig:lambda-functions-fxr}), at the expense of wider energy gaps between ordered and disordered states. With any of the soft-core settings, the $\lambda$-functions for the FXR/26 system we obtained indicate that the RatSC soft-core function to be significantly superior to the TanhSC soft-core functions in reducing the binding energy gaps between ordered and disordered states and lead to improved RE efficiency.

\subsection{Design of the Integrated Logistic Alchemical Schedules\label{sec:ilogistic-schedule-design}}

The $\lambda$-functions obtained above are used to design alchemical schedules based on the integrated logistic perturbation function from Eq.~(\ref{eq:ilog-function}) to attempt to avoid order/disorder transitions or at least reduce their effects.\cite{Kim2010b} The main design principle (see Figure \ref{fig:lambdaf-t4l}) is to vary the $\lambda$ dependence of the parameters of the integrated logistic function so as that its derivative with respect to $u$ has a single intersection with the $\lambda$-function--yielding a binding energy distribution with a single maximum--or, when this is not easily achievable, at least at nearby points, thereby yielding maxima and minima separated by small energy gaps. The parameters of the optimized integrated logistic schedule for the systems studied in this work are listed in Tables \ref{tab:ilog-t4l}, \ref{tab:ilog-fxr-umax50}, and \ref{tab:ilog-fxr-umax100}.

\begin{figure}
  \includegraphics{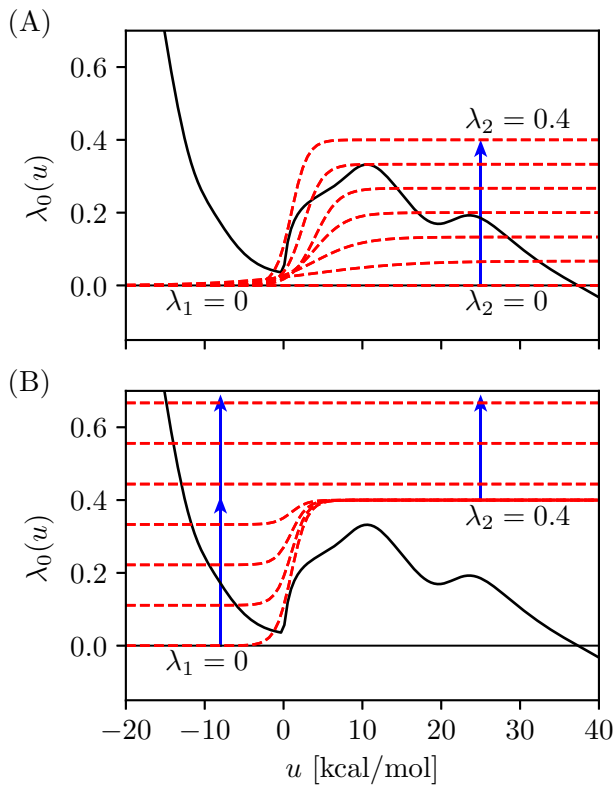}
  \caption{\label{fig:ilogistic-design} The general design strategy of the alchemical coupling schedule using the integrated logistic perturbation potential to avoid strong order/disorder transitions. Starting at the uncoupled ensemble with $\lambda_1 = 0$ and $\lambda_2 = 0$, first (A) $\lambda_2$ is increased up to a critical value (here $\lambda_c = 0.4$) sufficiently large to clear the maximum of $\lambda_0(u)$. Then (B) $\lambda_1$ is increased leaving $\lambda_2$ at the critical value $\lambda_c$. When $\lambda_1$ exceeds $\lambda_c$, $\lambda_2$ and $\lambda_1$ increase in unison, thereby restoring the linear alchemical schedule. The $\alpha$ and $u_0$ parameters which control the sharpness and location of the transition from $\lambda_1$ to $\lambda_2$, are adjusted slightly to ensure that the logistic function crosses the $\lambda$-function at only one point or at a set of points near each other. The format of the curves is the same as in Figure \ref{fig:lambdaf-t4l}A.}
\end{figure}

\begin{figure*}
  \includegraphics{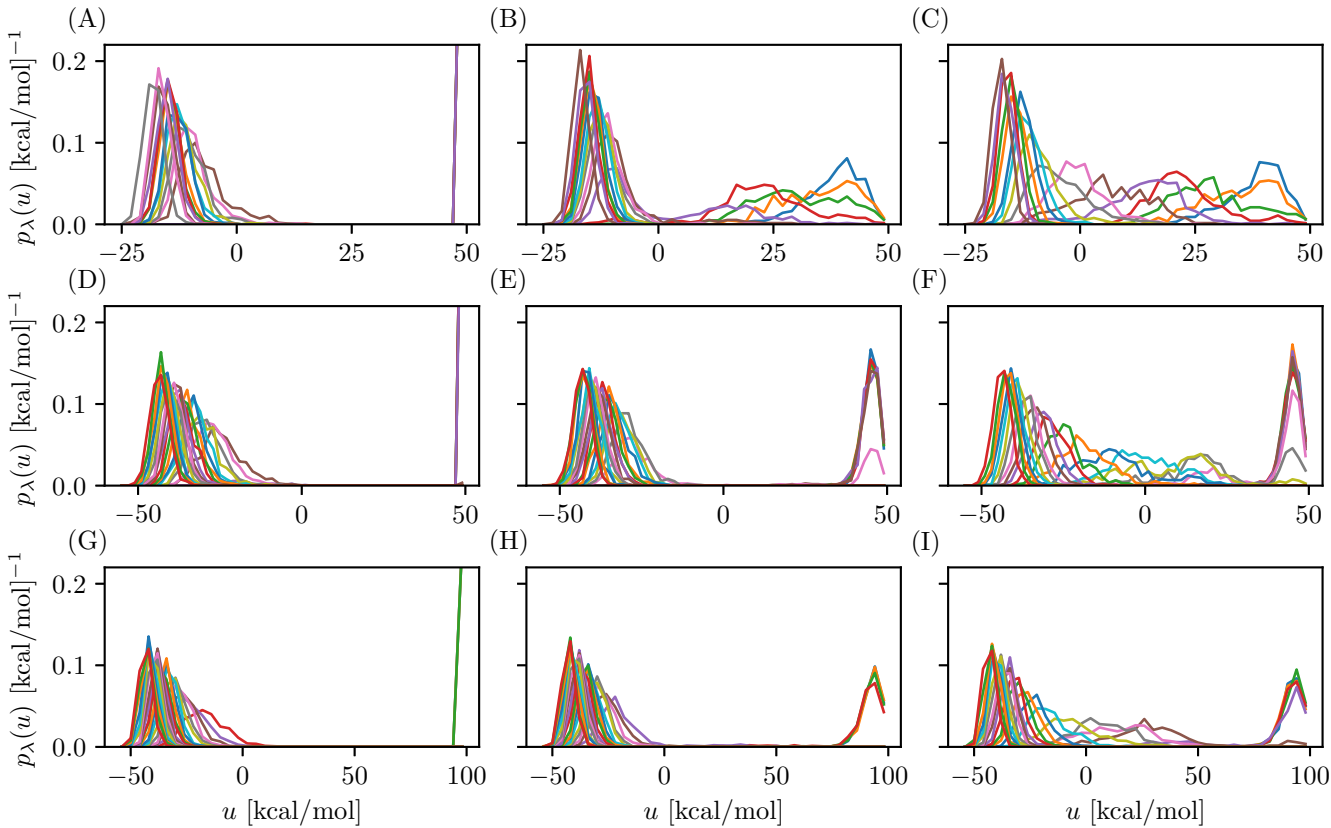}
  \caption{\label{fig:plambdas} The binding energy probability densities, $p_\lambda(u)$, collected from the Hamiltonian replica exchange simulations of T4L/3-iodotoluene (first row; panels A, B, C), FXR/26 with $u_{\rm max} = 50$ kcal/mol (second row; panels D, E, and F), and FXR/26 with $u_{\rm max} = 100$ kcal/mol (third row; panels G, H, and I). The first column of panels is with the TanhSC soft-core function and the linear alchemical perturbation. The second column is with the RatSC soft-core function and the linear alchemical perturbation. The third column is with the RatSC soft-core function and the integrated logistic perturbation. The colors of the distributions, which are repeated, correspond to different values of $\lambda$. The distributions shift left towards lower binding energies as $\lambda$ increases.  }
\end{figure*}

The general design strategy we followed in this work to parameterize the integrated logistic schedule is illustrated in Figure \ref{fig:ilogistic-design}. The integrated logistic potential essentially allows using different $\lambda$ values in different ranges of the binding energy. The coupling transformation, which with the linear alchemical potential involves the progressive increase of the single coupling parameter $\lambda$, is divided into three phases. In the first phase (Figure \ref{fig:ilogistic-design}A), the coupling parameter $\lambda_2$, for high binding energies is increased up to a critical value large enough to clear the maximum of the $\lambda$ function while $\lambda_1$, the coupling parameter for low binding energies, is left at zero. In the second phase (Figure \ref{fig:ilogistic-design}B) $\lambda_1$ is now increased while $\lambda_2$ is left unchanged at the critical value. The second phase ends when $\lambda_1$ reaches the critical value. Finally, in the third phase (Figure \ref{fig:ilogistic-design}B) $\lambda_1$ and $\lambda_2$ are increased in unison thereby acting as a linear perturbation up to $\lambda=1$. During the whole alchemical process $\partial W_\lambda(u)/\partial u$ intersects the $\lambda$-function at a single point, thereby yielding binding energy distributions with a single mode which progressively shifts to low binding energies without undergoing strong order/disorder transitions (Figure \ref{fig:plambdas}).

We followed the process outlined above to derive from the analytical $\lambda$-functions of the T4L/3-iodotoluene and FXR/26 systems (Figures \ref{fig:lambda-functions-t4l} and \ref{fig:lambda-functions-fxr}) optimized integrated logistic schedules capable of avoiding the phase transition at $u \simeq 0$ (Tables \ref{tab:ilog-t4l}, \ref{tab:ilog-fxr-umax50}, \ref{tab:ilog-fxr-umax100}). We have not attempted to design integrated logistic schedules aimed at resolving the phase transition of the FXR/26 system at large binding energies.

The binding energy probability distributions obtained from the molecular dynamics replica exchange calculations reported in Figures \ref{fig:plambdas}, largely confirm the predictions of the analytical model. With the TanhSC soft-core potential (panels A, B, and C), the binding energies of disordered states at $u \simeq u_{\rm max}$ abruptly shifts to negative values as $\lambda$ is lowered below a critical value ($\lambda = 0.2$ approximately for both systems and settings considered). The distributions of disordered and ordered states on either side of this critical value are separated by a large binding energy gap which increases from $50$ to $100$ kcal/mol as $u_{\rm max}$ is increased.

Consistent with the predicted $\lambda$-functions (Figures \ref{fig:lambda-functions-t4l} and \ref{fig:lambda-functions-fxr})), the binding energy gap is reduced when using the RatSC soft-core potential (Figure \ref{fig:plambdas}, panels D, E, and F), mainly by shifting the distributions at $u = u_{\rm max}$ to lower values. In the case of T4L/3-iodotoluene the effect is very substantial to the point that some of the distributions of the disordered states overlap, albeit weakly, with those of ordered states. As shown in Figure \ref{fig:lambdaf-t4l} the distributions near the critical $\lambda$ are bimodal. The benefit of the RatSC function is not as significant for the FXR/26 system, which is characterized by a stronger order/disorder transition.

The integrated logistic alchemical perturbation potential is very effective at canceling the transition near $u = 0$ (Figure \ref{fig:plambdas}, panels G, H, and I), especially for the T4L/3-iodotoluene system where it produces nearly homogeneous sampling of the whole binding energy range. The binding energy gap across the order/disorder transition For FXR/26 at large binding energies is reduced by a factor of five when using $u_{\rm max} = 50$ kcal/mol (panel F) and by a factor of two with $u_{\rm max} = 100$ kcal/mol (panel I). This result is due to the shifting of the distributions of the ordered state into the ``no man's land'' region of binding energies between ordered and disordered states which are very poorly sampled with the linear potential.

\subsection{Replica Exchange Efficiency\label{sec:replica-exchange-efficiency}}

\begin{figure*}
  \includegraphics{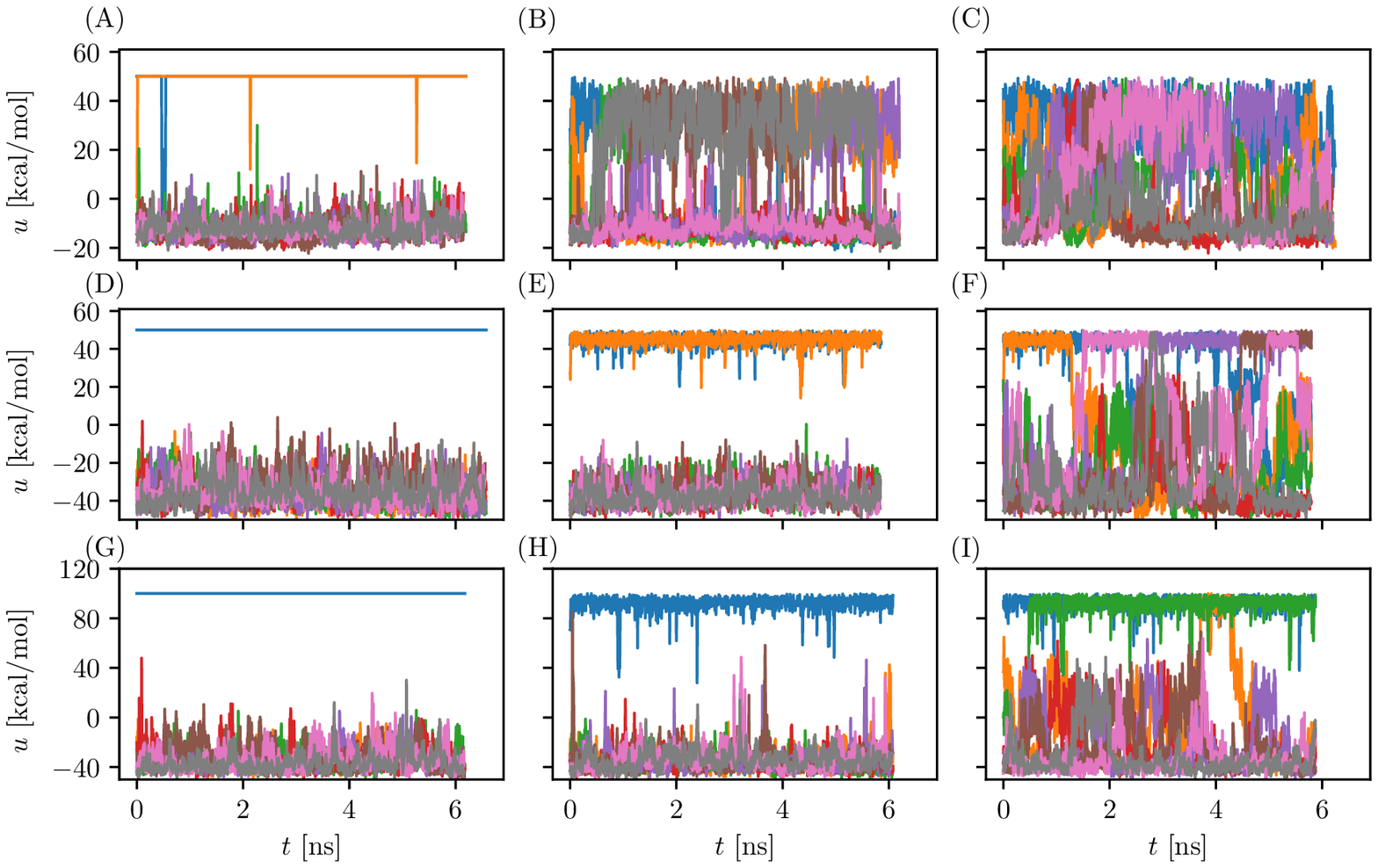}
  \caption{\label{fig:binde_traj} Binding energy trajectories for selected replicas from the Hamiltonian replica exchange simulations. The grid of panels follows the same order as in Figure \ref{fig:plambdas}. For T4L/3-iodotoluene (first row) only the odd-numbered replicas are shown (8 out of 16). For FXR/26 (second and third row) only one every 3 replicas are shown (8 out of 24). The dashed horizontal lines mark the $u_\text{lower}$ and $u_\text{upper}$ binding energy values used for the binding/unbinding transition counts in Table \ref{tab:binding_unbinding}.  }
\end{figure*}

Replica exchange efficiency has been monitored in terms of the extent of diffusion of replicas in $\lambda$-space and binding energy space. The time trajectories of the binding energies sampled by the replicas are shown in Figure \ref{fig:binde_traj}. They indicate that, while replicas rapidly equilibrate within the ordered low energy states and disordered high energy states of the complex, they more rarely interconvert between these states. We refer to these rare interconversions as binding or unbinding events. With the TanhSC soft-core potential replicas never, or only very rarely, undergo binding or unbinding transitions. The RatSC soft-core reduces the binding energy gap between ordered and disordered states and allowing the T4L/3-iodotoluene system to undergo numerous binding/unbinding transitions. However, only when also using the integrated logistic potential it is possible to observe binding/unbinding transitions for the FXR/26 system, which has a stronger order/disorder phase separation.

The number of binding and unbinding transitions, defined as the number of times that a replica goes from a disordered state to an ordered state or viceversa, are reported in Table \ref{tab:binding_unbinding}. Here we identify ordered and disordered states based on their binding energy values. If $u > u_{\rm upper}$ we label the replica as disordered (unbound) and if $u < u_{\rm lower}$ we label it as ordered (bound). The values for $u_{\rm upper}$ and $u_{\rm lower}$ (see Table \ref{tab:binding_unbinding}) were chosen based on the inspection of molecular dynamics trajectories. The trends are clear, confirm the qualitative analysis above, and do not depend on the specific choice of these parameters. In all cases, the TanhSC soft-core potential and the linear alchemical schedule lead to very few binding and unbinding transitions. As further discussed below, because of the lack of binding/unbinding events, the convergence of the calculations with the TanhSC soft-core potential is likely unreliable. The RatSC soft-core alone is sufficient to enable many interconversions for the T4L/3-iodotoluene system, one third more, in fact, than with the integrated logistic potential. The integrated logistic alchemical potential (ILog) produces many binding/unbinding events even for the more challenging FXR/26 system where the linear potential fails (Table \ref{tab:binding_unbinding}). Based on this analysis, the combination of the RatSC soft-core and the integrated logistic potential emerges as the most consistent scheme to accelerate binding/unbinding transitions.

\begin{table}
\caption{\label{tab:binding_unbinding} Number of binding and unbinding transitions }
\begin{ruledtabular}
\begin{tabular}{lcc}
  Protocol & $n_\text{bind}$ & $n_\text{unbind}$ \tabularnewline
\multicolumn{3}{c}{T4L99A/3-iodotoluene\footnote{$u_\text{lower} = -10$, $u_\text{upper} = 25$ kcal/mol, $t > 50$ ps}}\tabularnewline
TanhSC     &  $2$         & $2$   \tabularnewline
RatSC      &  $52$        & $55$  \tabularnewline
ILog       &  $29$        & $29$  \tabularnewline
\multicolumn{3}{c}{FXR/26 ($u_{\rm max} = 50$ kcal/mol)\footnote{$u_\text{lower} = -10$, $u_\text{upper} = 25$ kcal/mol, $t > 50$ ps}}\tabularnewline
TanhSC     &  $0$         & $1$  \tabularnewline
RatSC      &  $0$         & $1$  \tabularnewline
ILog       &  $41$        & $40$  \tabularnewline
\multicolumn{3}{c}{FXR/26 ($u_{\rm max}=100$ kcal/mol)\footnote{$u_\text{lower} = -20$, $u_\text{upper} = 50$ kcal/mol, $t > 50$ ps}}\tabularnewline
TanhSC     &  $0$         & $1$   \tabularnewline
RatSC      &  $2$         & $4$  \tabularnewline
ILog       &  $22$        & $22$  \tabularnewline
\end{tabular} 
\end{ruledtabular}  
\end{table}

\subsection{Binding Free Energy Estimates\label{sec:free-energy-results}}

\begin{figure*}
  \includegraphics{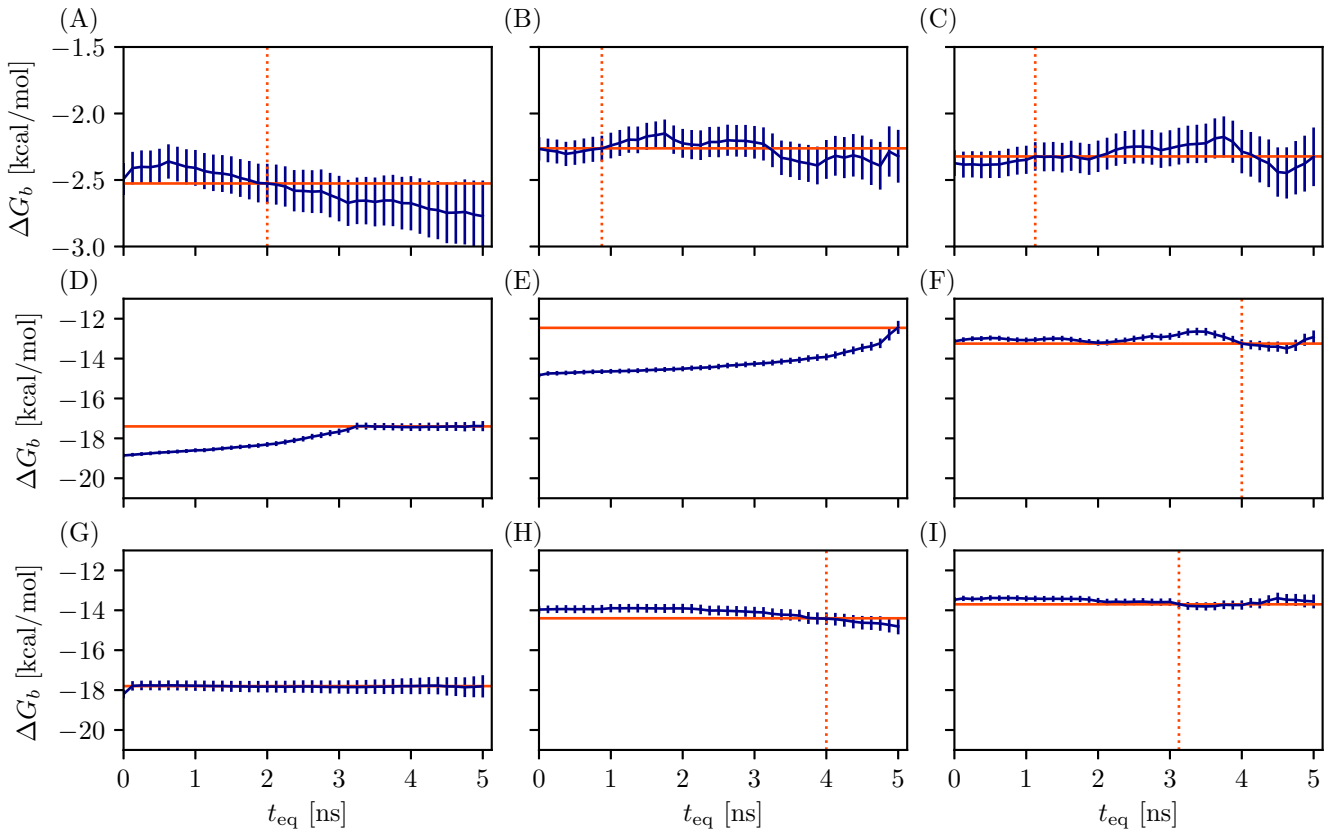}
  \caption{\label{fig:equilibration} Reverse cumulative profiles of the binding free energy for the complexes studied in this work. The grid of panels follows the same order as in Figure \ref{fig:plambdas}. Error bars are drawn at the $2 \sigma$ level. The horizontal line represents the best guess estimate of the binding free energy. The vertical dotted line (when present) represent the equilibration time above which free energy estimates agree within statistical uncertainty.}
\end{figure*}

The estimates of the standard free energy of binding for the systems and settings studied in this work are listed in Table \ref{tab:free_energies}. As the free energy is a thermodynamic state function, binding free energies should not depend on simulation settings, such as the choice of the soft-core function and the alchemical schedule. The data in Table \ref{tab:free_energies} shows that all the simulation protocols yield consistent estimates for the T4L/3-iodotoluene complex, thereby validating the correctness of our implementation. This is particularly so for the simulations with the linear (RatSC) and integrated logistic (ILog) alchemical schedules (Figures \ref{fig:equilibration}B and \ref{fig:equilibration}C) that achieve rapid equilibration between coupled and uncoupled states (Figures \ref{fig:binde_traj}B and \ref{fig:binde_traj}C).

Equilibration analysis (Figure \ref{fig:equilibration}) indicates that it takes approximately 1 ns of simulation per replica to achieve equilibration for the T4L/3-iodotoluene system with the RatSC and ILog protocols (see below). Consistent with the smaller number of binding and unbinding events (Table \ref{tab:binding_unbinding}), it takes about twice as long to achieve equilibration with the TanhSC protocol, although the steady downward drift of the binding free energy is a source of concern in this case (Figure \ref{fig:equilibration}A).

In this work, equilibration analysis is based on reverse cumulative profiles of the binding free energy with respect to the equilibration time, that is as a function the amount of data discarded from the beginning of the simulation.\cite{yang2004free,chodera2016simple,kilburg2018assessment} Based on these (Figure \ref{fig:equilibration}), it is qualitatively clear that binding free energy estimates for T4L/3-iodotoluene are relatively robust.

It is less obvious to derive quantitatively converged and minimum variance estimates from reverse cumulative profiles. While the accuracy of the binding free energy improves as more unequilibrated samples are discarded, the precision of the estimate worsens as fewer samples are available. The trade-off between accuracy and precision is reflected in the steady increase in the size of the error bars in Figure \ref{fig:equilibration} as the equilibration time increases. Several strategies have been proposed to extract optimal equilibration times and estimates from reverse cumulative profiles.\cite{yang2004free,chodera2016simple,kilburg2018assessment} Here we take the simple approach of choosing the smallest equilibration time that gives a free energy estimate statistically indistinguishable from those at all subsequent equilibration times. Equilibration times chosen using this strategy are indicated by vertical lines in Figure \ref{fig:equilibration} and are reported in the third column of Table \ref{tab:free_energies}.

Equilibration analysis based on the reverse cumulative plots indicates that the ILog protocol achieves equilibration of the binding free energy of the FXR/26 complex after 3 to 4 ns of simulation per replica, and the bias of the binding free energy estimate is less than half a kilocalorie per mole throughout the run (Figures \ref{fig:equilibration}F and \ref{fig:equilibration}I, and Table \ref{tab:free_energies}). In addition, the two variations of the ILog protocol with two different $u_{\rm max}$ parameters yield similar results ($-13.2$ and $-13.7$ kcal/mol, within nearly statistical uncertainty). These results, together with the relatively large number of binding/unbinding transitions (Table \ref{tab:binding_unbinding}), confer high confidence in the binding free energy estimates for the FXR/26 complex with the ILog protocol.

In contrast, there appears to be an almost complete lack of equilibration of the binding free energy of FXR/26 with the linear perturbation and the TanhSC soft-core settings.  The corresponding estimates are significantly more negative than with the other protocols and display little change as the equilibration time is increased (Figures \ref{fig:equilibration}D and \ref{fig:equilibration}G). This behavior is caused by the lack of unbinding transitions (Figures \ref{fig:equilibration}D and \ref{fig:equilibration}G) and it is of particular concern because conventional analysis would erroneously conclude in this case that the binding free energy estimate is well converged.

Similar, but less severe, issues occur for FXR/26 with the linear perturbation potential and the RatSC soft-core.  With $u_{\rm max} = 50$, despite the lack of binding/unbinding transitions, the binding free energy appears to equilibrate to the correct value on a time-scale longer than 5 ns (Figure \ref{fig:equilibration}E) per replica. With $u_{\rm max} = 100$ the binding free energy is near the correct value at most equilibration times, but it also equilibrates slowly, and towards the end appears to start to steer in the wrong direction (Figure \ref{fig:equilibration}H). In both cases equilibration of the binding free energy estimate is uncertain. This is in contrast to the calculations with the integrated logistic potential which, from all accounts, appear equilibrated and converged (Figures \ref{fig:equilibration}F and \ref{fig:equilibration}I).

\begin{table}
\caption{\label{tab:free_energies} Standard binding free energy estimates}
\begin{ruledtabular}
\begin{tabular}{lcc}
Protocol & $\Delta G^\circ_\text{site}$ & $t_\text{eq}$ \tabularnewline
\multicolumn{3}{c}{T4L/3-iodotoluene}\tabularnewline
TanhSC      & $-2.53 \pm 0.140$             &  $2.0$             \tabularnewline
RatSC       & $-2.26 \pm 0.086$             &  $0.9$             \tabularnewline
ILog        & $-2.32 \pm 0.097$             &  $1.1$             \tabularnewline
\multicolumn{3}{c}{FXR/26 ($u_{\rm max} = 50$ kcal/mol) }\tabularnewline
TanhSC      & $-17.4 \pm 0.3$\footnote{from last 1 ns of 6 ns per replica}  &  --             \tabularnewline
RatSC       & $-12.4 \pm 0.3$$^a$           &  --             \tabularnewline
ILog        & $-13.2 \pm 0.2$               &  $4.0$             \tabularnewline
\multicolumn{3}{c}{FXR/26 ($u_{\rm max} = 100$ kcal/mol) }\tabularnewline
TanhSC      & $-17.8 \pm 0.3$$^a$           &  --             \tabularnewline
RatSC       & $-14.4 \pm 0.3$               &  $4.0$             \tabularnewline
ILog        & $-13.7 \pm 0.2$               &  $3.2$             \tabularnewline
\end{tabular} 
\end{ruledtabular}  
\end{table}

\section{\label{sec:discussion} Discussion}


In this work, we show that order/disorder phase transition can occur along the alchemical transformations and that these, by impeding rapid interconversions between conformational states across the thermodynamic phase boundary, are responsible for slow equilibration of binding free energy estimates. The bottlenecks caused by order/disorder transitions are particularly noticeable in Hamiltonian Replica Exchange sampling approaches, where the perturbation energy gap causes reduces the rates of acceptance of $\lambda$-exchanges among replicas.

Many methods are available to activate interconversions between low energy basins separated by high energy barriers,\cite{Okumura2010,Pierce2012,abel2017advancing,xia2018improving} especially when the slow collective variable is known.\cite{ciccotti2004blue,barducci2008well,bussi2018metadynamics} Strategies to accelerate conformational sampling across entropic barriers, such as those considered here, are, however, not as established and understood. Probably the best known and still largely unsolved biophysical problem in this category is the process of protein folding, which is characterized by an entropic bottleneck to go from the disordered random coil state to the more ordered molten globule state.\cite{AlexeiV.Finkelstein2002,Zheng2007} Similarly, the process of protein-ligand binding involves loss of rotational and translational motion and the loss of conformational freedom of both the ligand and the receptor. In alchemical binding, this is reflected in the slow random search process of the rare ordered binding pose in the uncoupled ensemble, where the ligand and receptor are free to explore a wide range of configurations.\cite{Baron2008a,pan2017quantitative}

Here, we extend the formalism and non-Boltzmann conformational sampling simulation techniques developed by Straub at al.\ for the study of temperature-dependent phase transitions in spin systems, atomic clusters, and molecular liquids, to alchemical binding processes. There is a close analogy between the treatment of Straub et al.\cite{Kim2010b,kim2010generalized,lu2013order,lu2014investigating} for the canonical ensemble as a function of temperature and the alchemical ensemble as a function $\lambda$ considered here. Straub et al.\ considered non-Boltzmann sampling with a generalized reduced potential $W_T(E)$ where the modes of the energy distribution correspond to energies at which the inverse of the microcanonical statistical temperature function $T_S(E)$ equals the inverse of $k_B \partial W_T(E)/\partial E$:
\begin{equation}
  \frac{1}{T_S(E)} \equiv \frac{\partial S(E)}{\partial E} =
  k_B \frac{\partial W_T(E)}{\partial E} \, ,
  \label{eq:straub_condition}
\end{equation}
where $S(E) = k_B \ln\Omega(E)$ is the microcanonical entropy and $\Omega(E)$ is the density of states. In the case of canonical Boltzmann weighting, $W_T(E) = E/k_B T$, the condition Eq.~(\ref{eq:straub_condition}) reduces to $1/T_S(E) = 1/T = {\rm constant}$. Near a first order phase transition the function $1/T_S(E)$ varies non-monotonically and intersects the constant line $1/T$ at three energy values corresponding to the phases in equilibrium and the unstable intermediate phase. In close analogy with Figure (\ref{fig:phase-transition}), the phase transition can be avoided by using a suitable non-linear sampling potential $W_T(E)$ whose derivative intersects $1/T_S(E)$ at only one value of $E$.\cite{Kim2010b}

One of the main results of this work is the finding that the formalism of Straub et al.\ for temperature-driven order/disorder transitions extends nearly seamlessly to the alchemical ensemble for binding\cite{Gallicchio2010,Gallicchio2011adv} by applying the following equivalences:
\begin{eqnarray}
  E   & \rightarrow & u       \\
  \Omega(E) & \rightarrow & p_0(u) \\
  \frac{1}{k_B T} & \rightarrow & \beta \lambda \\
  \frac{1}{k_B T_S(E)}  & \rightarrow & \beta \lambda_0(u) \\
  W_T(E) & \rightarrow & W_\lambda(u)
\end{eqnarray}
We exploited these equivalences to design alchemical perturbation functions that avoid or reduce order/disorder phase transitions along the alchemical path. Using this strategies, we identified a particularly effective alchemical perturbation function in this respect, named the integrated logistic function [Eq.~(\ref{eq:ilog-function})], which we successfully employed to accelerate the sampling of two protein-ligand complexes affected by order/disorder transitions.

A critical step in achieving this result has been the description in analytical form of the $p_0(u)$ probability distribution function and the corresponding $\lambda$-function [Eq.~(\ref{eq:lambda-function})].\cite{kilburg2018analytical} Estimation of the $\lambda$-function by numerical histogramming and finite difference techniques, in particular, leads to noisy results which are hard to interpret and utilize for the design of optimal perturbation functions. In contrast, analytical differentiation of $p_0(u)$ has proven a suitable strategy to obtain $\lambda_0(u)$ and formulate reliable predictions on the strength of order/disorder transitions, and on how to avoid them in numerical simulations.

The analytical route relies on a statistical theory of alchemical binding developed earlier.\cite{kilburg2018analytical} The theory is based on a small number of physically-motivated parameters for each binding pose. The parameters are optimized by fitting the analytical predictions to binding energy distributions extracted from numerical simulations. Once the parameters are determined with some confidence, the theory is used to explore the effect of varying simulation conditions, such as alternative alchemical perturbation potentials and soft-core functions, as in the present work.  At the core of the theory is the assumption that the binding energy is a random variable with two components, one that follows the central limit statistics and the other following max statistics\cite{GumbelBook} describing, respectively, favorable ``soft'' interatomic interactions and atomic clashes. Here the theory is used for the first time to describe order/disorder transitions in alchemical calculations.

\section{Conclusions}

In this work, we investigate the role of order/disorder transitions in alchemical simulations of protein-ligand binding free energies. We show for a benchmarking system (the complex between the L99A mutant of T4 lysozyme and 3-iodotoluene) and for a more challenging system relevant for medicinal applications (the complex of the farnesoid X receptor and inhibitor 26 from a recent D3R challenge) that order/disorder transitions can significantly hamper Hamiltonian replica exchange sampling efficiency and slow down the rate of equilibration of binding free energy estimates. We further show that our analytical model of alchemical binding\cite{kilburg2018analytical} combined with the formalism developed by Straub et al.\ for the treatment of order/disorder transitions of molecular systems\cite{Kim2010b} can be successfully employed to analyze the transitions and help design alchemical schedules and soft-core functions that avoid or reduce the asdverse effects of rare binding/unbinding transitions. Future work from our laboratory will explore the use of these techniques for alchemical free energy estimation with explicit solvation, specifically hydration free energies and double-decoupling binding free energies, for systems undergoing order/disorder transitions.

\begin{acknowledgments}
We acknowledge support from the National Science Foundation (NSF
CAREER 1750511 to E.G.). Molecular simulations were conducted on the
Comet GPU supercomputer cluster at the San Diego Supercomputing Center
supported by NSF XSEDE award TG-MCB150001 and on a GPU cluster at the
Laboratory of Computational Molecular Biophysics at Brooklyn College
of CUNY.
\end{acknowledgments}

\appendix

\section{Appendix}

\subsection{\label{sec:a-pc}Derivation of the Pair Collisional Distribution Function}

\begin{figure}
  \includegraphics[scale=0.5]{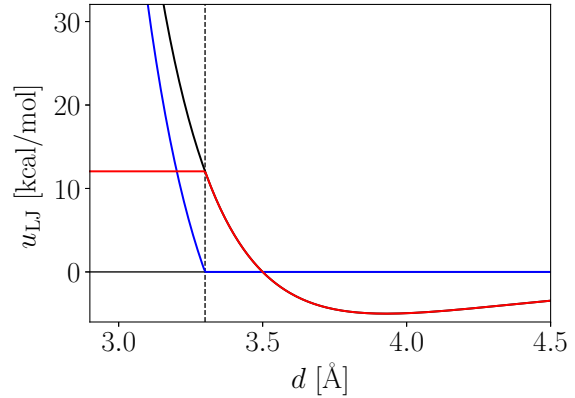}
\caption{\label{fig:wca_pot}Decomposition at the cutoff distance $r_{C}$ (dashed line) of the
Lennard-Jones potential (black) into repulsive (blue) and background
(red) components {[}see Eqs.~(\ref{eq:WCApot}) and (\ref{eq:WCApot-1}){]}}.
\end{figure}

Consider the decomposition of the Lennard-Jones pair potential $u_{{\rm LJ}}(r)=4\epsilon[(\sigma/r)^{12}-(\sigma/r)^{6}]$
into repulsive and background components based on the cutoff distance
$r_{C}$ (Fig.~\ref{fig:wca_pot}).
\begin{equation}
u_{C}(r)=\begin{cases}
u_{{\rm LJ}}(r)-\tilde{u}, & r<r_{C}\\
0 & r\ge r_{C}\:,
\end{cases}\label{eq:WCApot}
\end{equation}
\begin{equation}
u_{B}(r)=\begin{cases}
\tilde{u}, & r<r_{C}\\
u_{{\rm LJ}}(r) & r\ge r_{C}\:,
\end{cases}\label{eq:WCApot-1}
\end{equation}
where
\begin{equation}
\tilde{u}=u_{LJ}(r_{C}) > -\epsilon \label{eq:cutoff-energy}
\end{equation}
is the value of the LJ potential at the cutoff distance $r_{C}$.
The standard WCA decomposition of the LJ potential is a particular
case of Eqs.~(\ref{eq:WCApot}) and (\ref{eq:WCApot-1}) with a
cutoff distance corresponding to the minimum of the LJ potential,
$r_{C}=2^{1/6}\sigma$.

In this work, we model the collisional and background components
of the binding energy between the receptor and the ligand in the uncoupled
state are due to the repulsive and background component of the LJ
potential, Eqs.~(\ref{eq:WCApot}) and (\ref{eq:WCApot-1}) respectively,
for some choice of the LJ cutoff energy $\tilde{u}$ {[}Eq.~(\ref{eq:cutoff-energy}){]}.
This choice is justified as follows. The collisional component is
defined as the contribution to the binding energy described by max
statistics.\cite{kilburg2018analytical} Max statistics applies when the interaction energy is
dominated by the single pair of receptor and ligand atoms with the
largest repulsive interaction energy. Thus, max statistics can set
in only if the repulsive interaction energy potential, as in Eq.~(\ref{eq:WCApot}),
grows rapidly for small interatomic distances, which are less likely. Conversely, the collisional energy should become negligibly small for large interatomic distances, which
are more likely and tend to occur for many pairs of atoms at once.
In contrast, the background component is defined as the contribution
to the binding energy described by central limit statistics. Because
central limit statistics assumes contributions from many interatomic
interaction energies of similar magnitude and likelihood of occurrence,
it cannot apply to a potential which grows indefinitively with decreasing
interatomic distance. A sensible choice for the background component
is therefore an interatomic potential that is constant below a certain
interatomic distance, as in Eq.~(\ref{eq:WCApot-1}), thereby leaving
its complement, Eq.~(\ref{eq:WCApot}), as the necessary description
for the repulsive component of the LJ potential. 

The choice of Eq.~(\ref{eq:WCApot}) as a viable description of the
repulsive component of the interatomic potential is further supported
by the requirement that it should grow sufficiently rapidly with decreasing
interatomic distance such that the average and variance of the repulsive
pairwise energy tend to infinity, so as to exclude the applicability
of central-limit statistics to the collisional component. The probability
distribution of the collisional component derived from Eq.~(\ref{eq:WCApot})
satisfy this requirement.

Finally, our collisional interatomic potential energy model does not
include electrostatic interactions. This is partly because of the
difficulty of representing the statistics of contributions of alternating
sign coming from like and unlike partial charges. Nevertheless, we
expect our LJ-based statistical model to be accurate at short interatomic
distances where the $1/r^{12}$ repulsion dominates over the much
weaker $1/r$ behavior of the electrostatic interactions. The contributions
of electrostatic interactions to the background component of the binding
energy are fully captured by the average background interaction energy
parameter $\bar{u}_{B}$ (see above). 

Now consider two particles interacting by the repulsive LJ potential
(\ref{eq:WCApot}) in which one particle (representing the receptor)
is fixed at the origin and the other (representing the ligand) is
uniformly distributed in a sphere of radius $r_{C}$ centered at the
origin (Fig.~\ref{fig:wca_pot}). We will derive the probability
density $p_{C}(u_{C})$ of the repulsive interaction energy $u_{{\rm C}}(r)$,
where $r$ is the distance between the two particles, by differentiating
the cumulative probability function $P_{C}(u_{C})$ defined as the
probability that, given that the ligand particle is uniformly distributed
in the sphere, the interaction energy $u_{{\rm C}}(r)$ is greater
than the given value $u_{C}$. 

The probability that the pair interaction energy is smaller than $u_{C}$
is given by:
\begin{equation}
P_{C}(u_{C})=H(u_{C})\frac{V_{C}-V(u_{C})}{V_{C}}\label{eq:pcu_ratio-1}
\end{equation}
where the Heaviside function imposes the requirement that $u_{C}$
be larger than zero, $V_{C}$ is the volume of the sphere of radius
$r_{c}$ and $V(u_{C})$ is the volume of the sphere of radius $r(u_{C})$,
where $r(u_{C})$ is inter-particle distance at which the repulsive
LJ potential has value $u_{C}$. From Eq.~(\ref{eq:WCApot}) we have
\begin{equation}
r(u_{C})=\frac{r_{0}}{(1+x_{C})^{1/6}};\quad u_{C}\ge0\label{eq:ru-2}
\end{equation}
where $r_{0}=2^{1/6}\sigma_{LJ}$ is the minimum of the Lennard-Jones
pair potential and
\begin{equation}
x_{C}=\sqrt{\frac{u_{C}}{\epsilon}+\frac{\tilde{u}}{\epsilon}+1}\label{eq:x-def}
\end{equation}
Inserting Eq.~(\ref{eq:ru-2}) into Eq.~(\ref{eq:pcu_ratio-1})
gives
\begin{equation}
P_{C}(u_{C})=H(u_{C})\frac{V_{C}-V(u_{C})}{V_{C}}\label{eq:pcu_ratio-1-1}
\end{equation}
Differentiating \ref{eq:pcu_ratio-1-1} with respect to $u_{C}$ and
setting 
\begin{equation}
\tilde{x}_{C}=\sqrt{\frac{\tilde{u}}{\epsilon}+1}
\end{equation}
finally yields
\begin{equation}
p_C(u) = \frac{H(u_{C})(1+\tilde{x}_{C})^{1/2}}{4\epsilon x_C(1+x_C)^{3/2}}\label{eq:PC(u)} \, ,
\end{equation}
which expresses a normalized
distribution as it can be verified by direct integration using the
fact that 
\begin{equation}
\int\frac{dx}{(1+x)^{3/2}}=-\frac{2}{(1+x)^{1/2}}\label{eq:ru-1-1}
\end{equation}
Eq.~(\ref{eq:PC(u)}) is the basis of the derivation of the collisional energy distribution function in Eq.~(\ref{eq:pwcaf}) for a polyatomic ligand interacting with a polyatomic receptor as discussed in reference \onlinecite{kilburg2018analytical}.

\begin{table}
  \caption{\label{tab:ilog-t4l} Alchemical schedule of the integrated logistic perturbation function for the T4L/3-iodotoluene complex}
\begin{ruledtabular}
\begin{tabular}{lccccc}
$\lambda$  &   $\lambda_1$ & $\lambda_2$ & $\alpha$\footnote{kcal/mol$^{-1}$} & $u_0$\footnote{kcal/mol} & $w_0$\footnote{kcal/mol}  \tabularnewline
0.000 & 0.000 & 0.000 & 0.010 & 4.000 & 0.000 \tabularnewline
0.067 & 0.000 & 0.067 & 0.167 & 4.000 & -0.544 \tabularnewline
0.133 & 0.000 & 0.133 & 0.333 & 4.000 & -0.811 \tabularnewline
0.200 & 0.000 & 0.200 & 0.500 & 4.000 & -1.077 \tabularnewline
0.267 & 0.000 & 0.267 & 0.667 & 4.000 & -1.344 \tabularnewline
0.333 & 0.000 & 0.333 & 0.833 & 2.504 & -1.112 \tabularnewline
0.400 & 0.000 & 0.400 & 1.000 & 1.000 & -0.677 \tabularnewline
0.467 & 0.111 & 0.400 & 1.000 & 1.000 & -0.289 \tabularnewline
0.533 & 0.222 & 0.400 & 1.000 & 1.000 & -0.178 \tabularnewline
0.600 & 0.333 & 0.400 & 1.000 & 1.000 & -0.067 \tabularnewline
0.667 & 0.444 & 0.444 & 1.000 & 1.000 & 0.000 \tabularnewline
0.733 & 0.556 & 0.556 & 1.000 & 1.000 & -0.000 \tabularnewline
0.800 & 0.667 & 0.667 & 1.000 & 1.000 & 0.000 \tabularnewline
0.867 & 0.778 & 0.778 & 1.000 & 1.000 & 0.000 \tabularnewline
0.933 & 0.889 & 0.889 & 1.000 & 1.000 & 0.000 \tabularnewline
1.000 & 1.000 & 1.000 & 1.000 & 1.000 & 0.000 \tabularnewline
\end{tabular} 
\end{ruledtabular}
\end{table}

\begin{table}
  \caption{\label{tab:ilog-fxr-umax50} Alchemical schedule of the integrated logistic perturbation function for the FXR/26 complex with $u_{\rm max} = 50$ kcal/mol}
\begin{ruledtabular}
\begin{tabular}{lccccc}
$\lambda$  &   $\lambda_1$ & $\lambda_2$ & $\alpha$\footnote{kcal/mol$^{-1}$} & $u_0$\footnote{kcal/mol} & $w_0$\footnote{kcal/mol}  \tabularnewline
0.000 & 0.000 & 0.000 & 0.350 & 2.000 & 0.000 \tabularnewline
0.043 & 0.000 & 0.056 & 0.350 & 2.000 & -0.223 \tabularnewline
0.087 & 0.000 & 0.111 & 0.350 & 2.000 & -0.442 \tabularnewline
0.130 & 0.000 & 0.167 & 0.350 & 2.000 & -0.665 \tabularnewline
0.174 & 0.000 & 0.222 & 0.350 & 2.000 & -0.884 \tabularnewline
0.217 & 0.000 & 0.278 & 0.350 & 2.000 & -1.107 \tabularnewline
0.261 & 0.000 & 0.334 & 0.350 & 2.000 & -1.329 \tabularnewline
0.304 & 0.000 & 0.389 & 0.350 & 2.000 & -1.548 \tabularnewline
0.348 & 0.000 & 0.445 & 0.350 & 2.000 & -1.771 \tabularnewline
0.391 & 0.000 & 0.500 & 0.350 & 2.000 & -1.990 \tabularnewline
0.435 & 0.072 & 0.500 & 0.350 & 2.000 & -1.848 \tabularnewline
0.478 & 0.143 & 0.500 & 0.350 & 2.000 & -1.707 \tabularnewline
0.522 & 0.215 & 0.500 & 0.350 & 1.500 & -1.314 \tabularnewline
0.565 & 0.286 & 0.500 & 0.350 & 1.000 & -0.924 \tabularnewline
0.609 & 0.357 & 0.500 & 0.350 & 0.500 & -0.533 \tabularnewline
0.652 & 0.429 & 0.500 & 0.350 & 0.000 & -0.141 \tabularnewline
0.696 & 0.500 & 0.500 & 0.350 & 0.000 & 0.000 \tabularnewline
0.739 & 0.572 & 0.572 & 0.350 & 0.000 & 0.000 \tabularnewline
0.783 & 0.643 & 0.643 & 0.350 & 0.000 & 0.000 \tabularnewline
0.826 & 0.714 & 0.714 & 0.350 & 0.000 & 0.000 \tabularnewline
0.870 & 0.786 & 0.786 & 0.350 & 0.000 & 0.000 \tabularnewline
0.913 & 0.857 & 0.857 & 0.350 & 0.000 & 0.000 \tabularnewline
0.957 & 0.929 & 0.929 & 0.350 & 0.000 & 0.000 \tabularnewline
1.000 & 1.000 & 1.000 & 0.350 & 0.000 & 0.000 \tabularnewline
\end{tabular} 
\end{ruledtabular}
\end{table}

\begin{table}
  \caption{\label{tab:ilog-fxr-umax100} Alchemical schedule of the integrated logistic perturbation function for the FXR/26 complex with $u_{\rm max} = 100$ kcal/mol}
\begin{ruledtabular}
\begin{tabular}{lccccc}
$\lambda$  &   $\lambda_1$ & $\lambda_2$ & $\alpha$\footnote{kcal/mol$^{-1}$} & $u_0$\footnote{kcal/mol} & $w_0$\footnote{kcal/mol}  \tabularnewline
0.000 & 0.000 & 0.000 & 0.250 & 5.000 & 0.000 \tabularnewline
0.043 & 0.000 & 0.037 & 0.250 & 5.000 & -0.288 \tabularnewline
0.087 & 0.000 & 0.075 & 0.250 & 5.000 & -0.583 \tabularnewline
0.130 & 0.000 & 0.112 & 0.250 & 5.000 & -0.871 \tabularnewline
0.174 & 0.000 & 0.149 & 0.250 & 5.000 & -1.158 \tabularnewline
0.217 & 0.000 & 0.187 & 0.250 & 5.000 & -1.453 \tabularnewline
0.261 & 0.000 & 0.224 & 0.250 & 5.000 & -1.741 \tabularnewline
0.304 & 0.000 & 0.261 & 0.250 & 5.000 & -2.029 \tabularnewline
0.348 & 0.063 & 0.261 & 0.250 & 4.000 & -1.593 \tabularnewline
0.391 & 0.125 & 0.261 & 0.250 & 3.000 & -1.160 \tabularnewline
0.435 & 0.188 & 0.261 & 0.250 & 1.000 & -0.463 \tabularnewline
0.478 & 0.250 & 0.261 & 0.250 & 0.000 & -0.030 \tabularnewline
0.522 & 0.313 & 0.313 & 0.250 & 0.000 & 0.000 \tabularnewline
0.565 & 0.375 & 0.375 & 0.250 & 0.000 & 0.000 \tabularnewline
0.609 & 0.438 & 0.438 & 0.250 & 0.000 & 0.000 \tabularnewline
0.652 & 0.500 & 0.500 & 0.250 & 0.000 & 0.000 \tabularnewline
0.696 & 0.563 & 0.563 & 0.250 & 0.000 & 0.000 \tabularnewline
0.739 & 0.625 & 0.625 & 0.250 & 0.000 & 0.000 \tabularnewline
0.783 & 0.688 & 0.688 & 0.250 & 0.000 & 0.000 \tabularnewline
0.826 & 0.750 & 0.750 & 0.250 & 0.000 & 0.000 \tabularnewline
0.870 & 0.813 & 0.813 & 0.250 & 0.000 & 0.000 \tabularnewline
0.913 & 0.875 & 0.875 & 0.250 & 0.000 & 0.000 \tabularnewline
0.957 & 0.938 & 0.938 & 0.250 & 0.000 & 0.000 \tabularnewline
1.000 & 1.000 & 1.000 & 0.250 & 0.000 & 0.000 \tabularnewline
\end{tabular} 
\end{ruledtabular}
\end{table}

\subsection{\label{sec:gibbs} Gibbs Independence Sampling Algorithm}

While it not central to the present analysis of alchemically-induced phase transitions, for completeness, here we describe the implementation of the Gibbs Independence Sampling algorithm we used in this work in conjunction with asynchronous replica exchange. The scheme has been used previously,\cite{pal2016SAMPL5,bfzhang2016rnaseh} but has not yet been reported explicitly in the literature.

The space of assignments of $\lambda$-states to replicas is composed of $M!$ discrete members each corresponding to a permutation of $M$ replicas on $M$ $\lambda$-states. A permutation $r$ is defined by the set of indexes $\{ k_i(r) \}$ specifying the state $k$ assigned to each replica $i$.
In our specific implementation of Gibbs Independence Sampling, starting with current state assignment $r$, we consider transitioning to a set of $M-1$ new state permutations $s$ differing from $r$ by a single swap of $\lambda$-states between a chosen pivot replica $i$ and any other replica $j$. The discrete Metropolis transition matrix\cite{Allen:Tildesley} of this random walk in state permutations is
\begin{eqnarray}
  T_{rs} & = & \alpha_{rs} \min \left[ 1, e^{-( v_s - v_r  )}  \right] \nonumber \\
  T_{rr} & = & 1 - \sum_{s \ne r} T_{rs} \label{eq:Gibbs-Trs}
\end{eqnarray}
where the proposal probability $\alpha_{rs}$ is $1/(M-1)$ if the permutations differ by a single replica swap between replica $i$ and some other replica $j$ as described above and zero otherwise, and $v_{r}$ is the reduced generalized bias potential energy of the replica exchange ensemble corresponding to the state assignment $r$:
\begin{equation}
  v_r = \sum_{i} \beta W_{k_i(r)}(u_i)
\end{equation}
where $W_{k}(u_i)$ is the alchemical perturbation energy [Eq.~(\ref{eq:pert_pot})] of replica $i$ at $\lambda_k$.

Given a pivot replica $i$, the algorithm selects a replica $j$ for exchange with probability Eq.~(\ref{eq:Gibbs-Trs}). The selected replica could include the same replica $i$, in which case no exchange occurs. The process is repeated sequentially for each replica $i$ so that at the end of the exchange cycle each replica has had the opportunity to exchange with any other replica. It is straightforward to show the random walk produced by this algorithm satisfy microscopic reversibility. To reduce computational cost, the set of energies $W_{k}(u_i)$ is pre-computed at the beginning of each exchange cycle. We prefer this algorithm over alternatives,\cite{chodera2011replica} because it automatically selects likely exchanges in an unbiased manner without having to specify neighboring states. Moreover, it naturally scales appropriately the number of exchange moves in proportion to the number of replicas.


\begin{thebibliography}{10}

\bibitem{GallicchioSAMPL4}
Emilio Gallicchio, Nanjie Deng, Peng He, Alexander~L. Perryman, Daniel~N.
  Santiago, Stefano Forli, Arthur~J. Olson, and Ronald~M. Levy.
\newblock Virtual screening of integrase inhibitors by large scale binding free
  energy calculations: the {SAMPL4} challenge.
\newblock {\em J. Comp. Aided Mol. Des.}, 28:475--490, 2014.

\bibitem{gaieb2018d3r}
Zied Gaieb, Shuai Liu, Symon Gathiaka, Michael Chiu, Huanwang Yang, Chenghua
  Shao, Victoria~A Feher, W~Patrick Walters, Bernd Kuhn, Markus~G Rudolph,
  et~al.
\newblock D3r grand challenge 2: blind prediction of protein--ligand poses,
  affinity rankings, and relative binding free energies.
\newblock {\em J. Comp. Aid. Mol. Des.}, 32(1):1--20, 2018.

\bibitem{Mobley:Dill:review:2009}
David~L Mobley and Ken~A Dill.
\newblock Binding of small-molecule ligands to proteins: "what you see" is not
  always "what you get".
\newblock {\em Structure}, 17:489--498, 2009.

\bibitem{pal2016SAMPL5}
Rajat~Kumar Pal, Kamran Haider, Divya Kaur, William Flynn, Junchao Xia,
  Ronald~M. Levy, Tetiana Taran, Lauren Wickstrom, Tom Kurtzman, and Emilio
  Gallicchio.
\newblock A combined treatment of hydration and dynamical effects for the
  modeling of host-guest binding thermodynamics: The {SAMPL5} blinded
  challenge.
\newblock {\em J. Comp. Aided Mol. Des.}, 31:29--44, 2016.

\bibitem{henderson2018ligand}
Jack~A Henderson, Robert~C Harris, Cheng-Chieh Tsai, and Jana Shen.
\newblock How ligand protonation state controls water in protein--ligand
  binding.
\newblock {\em J. Phys. Chem. Lett.}, 9(18):5440--5444, 2018.

\bibitem{cruzeiro2018redox}
Vin{\'\i}cius Wilian~D Cruzeiro, Marcos~S Amaral, and Adrian~E Roitberg.
\newblock Redox potential replica exchange molecular dynamics at constant ph in
  amber: Implementation and validation.
\newblock {\em J. Chem. Phys.}, 149(7):072338, 2018.

\bibitem{harder2015opls3}
Edward Harder, Wolfgang Damm, Jon Maple, Chuanjie Wu, Mark Reboul, Jin~Yu
  Xiang, Lingle Wang, Dmitry Lupyan, Markus~K Dahlgren, Jennifer~L Knight,
  et~al.
\newblock Opls3: a force field providing broad coverage of drug-like small
  molecules and proteins.
\newblock {\em J. Chem. Theory Comput.}, 12(1):281--296, 2015.

\bibitem{Albaugh2016advanced}
Alex Albaugh, Henry~A. Boateng, Richard~T. Bradshaw, Omar~N. Demerdash, Jacek
  Dziedzic, Yuezhi Mao, Daniel~T. Margul, Jason Swails, Qiao Zeng, David~A.
  Case, Peter Eastman, Lee-Ping Wang, Jonathan~W. Essex, Martin Head-Gordon,
  Vijay~S. Pande, Jay~W. Ponder, Yihan Shao, Chris-Kriton Skylaris, Ilian~T.
  Todorov, Mark~E. Tuckerman, and Teresa Head-Gordon.
\newblock Advanced potential energy surfaces for molecular simulation.
\newblock {\em J. Phys. Chem. B}, 120(37):9811--9832, 2016.
\newblock PMID: 27513316.

\bibitem{wang2017building}
Lee-Ping Wang, Keri~A McKiernan, Joseph Gomes, Kyle~A Beauchamp, Teresa
  Head-Gordon, Julia~E Rice, William~C Swope, Todd~J Mart{\'\i}nez, and Vijay~S
  Pande.
\newblock Building a more predictive protein force field: a systematic and
  reproducible route to amber-fb15.
\newblock {\em J. Phys. Chem. B}, 121(16):4023--4039, 2017.

\bibitem{roos2019opls3e}
Katarina Roos, Chuanjie Wu, Wolfgang Damm, Mark Reboul, James~M Stevenson, Chao
  Lu, Markus~K Dahlgren, Sayan Mondal, Wei Chen, Lingle Wang, et~al.
\newblock Opls3e: Extending force field coverage for drug-like small molecules.
\newblock {\em Journal of chemical theory and computation}, 15(3):1863--1874,
  2019.

\bibitem{Lapelosa2011}
Mauro Lapelosa, Emilio Gallicchio, and Ronald~M. Levy.
\newblock Conformational transitions and convergence of absolute binding free
  energy calculations.
\newblock {\em J. Chem. Theory Comput.}, 8:47--60, 2012.

\bibitem{Mobley2012}
David~L Mobley.
\newblock Let's get honest about sampling.
\newblock {\em J Comput Aided Mol Des}, 26:93--95, 2012.

\bibitem{bodnarchuk2014strategies}
Michael~S Bodnarchuk, Russell Viner, Julien Michel, and Jonathan~W Essex.
\newblock Strategies to calculate water binding free energies in
  protein--ligand complexes.
\newblock {\em J. Chem. Inf. Mod.}, 54(6):1623--1633, 2014.

\bibitem{procacci2019solvation}
Piero Procacci.
\newblock Solvation free energies via alchemical simulations: let's get honest
  about sampling, once more.
\newblock {\em Phys. Chem. Chem. Phys.}, 2019.

\bibitem{Jorgensen2009}
William~L Jorgensen.
\newblock Efficient drug lead discovery and optimization.
\newblock {\em Acc Chem Res}, 42:724--733, 2009.

\bibitem{michel2010prediction}
Julien Michel and Jonathan~W Essex.
\newblock Prediction of protein--ligand binding affinity by free energy
  simulations: assumptions, pitfalls and expectations.
\newblock {\em J. Comp. Aid. Mol. Des.}, 24:639--658, 2010.

\bibitem{Gallicchio2011adv}
Emilio Gallicchio and Ronald~M Levy.
\newblock Recent theoretical and computational advances for modeling
  protein-ligand binding affinities.
\newblock {\em Adv. Prot. Chem. Struct. Biol.}, 85:27--80, 2011.

\bibitem{limongelli2013funnel}
Vittorio Limongelli, Massimiliano Bonomi, and Michele Parrinello.
\newblock Funnel metadynamics as accurate binding free-energy method.
\newblock {\em Proc. Natl. Acad. Sci.}, 110(16):6358--6363, 2013.

\bibitem{deng2018comparing}
Nanjie Deng, Di~Cui, Bin~W Zhang, Junchao Xia, Jeffrey Cruz, and Ronald Levy.
\newblock Comparing alchemical and physical pathway methods for computing the
  absolute binding free energy of charged ligands.
\newblock {\em Phys. Chem. Chem. Phys.}, 20(25):17081--17092, 2018.

\bibitem{lybrand:mccammon1986}
Terry~P Lybrand, J~Andrew McCammon, and Georges Wipff.
\newblock Theoretical calculation of relative binding affinity in host-guest
  systems.
\newblock {\em Proc. Natl. Acad. Sci. USA}, 83(4):833--835, 1986.

\bibitem{Shirts:Mobley:Chodera:2007:review}
M.R. Shirts, D.L. Mobley, and J.D. Chodera.
\newblock Alchemical free energy calculations: ready for prime time?
\newblock {\em Ann. Rep. Comput. Chem.}, 3:41--59, 2007.

\bibitem{Wang2012}
Lingle Wang, B.~J. Berne, and Richard~A. Friesner.
\newblock On achieving high accuracy and reliability in the calculation of
  relative protein-ligand binding affinities.
\newblock {\em Proc. Natl. Acad. Sci.}, 109:1937--1942, 2012.

\bibitem{homeyer2014binding}
Nadine Homeyer, Friederike Stoll, Alexander Hillisch, and Holger Gohlke.
\newblock Binding free energy calculations for lead optimization: assessment of
  their accuracy in an industrial drug design context.
\newblock {\em J. Chem. Theory Comput.}, 10(8):3331--3344, 2014.

\bibitem{abel2017advancing}
Robert Abel, Lingle Wang, Edward~D Harder, BJ~Berne, and Richard~A Friesner.
\newblock Advancing drug discovery through enhanced free energy calculations.
\newblock {\em Acc. Chem. Res.}, 50(7):1625--1632, 2017.

\bibitem{lee2018gpu}
Tai-Sung Lee, David~S Cerutti, Dan Mermelstein, Charles Lin, Scott LeGrand,
  Timothy~J Giese, Adrian Roitberg, David~A Case, Ross~C Walker, and Darrin~M
  York.
\newblock Gpu-accelerated molecular dynamics and free energy methods in
  amber18: performance enhancements and new features.
\newblock {\em J. Chem. Inf. Mod.}, 58(10):2043--2050, 2018.

\bibitem{Zou2019chemarxiv}
Junjie Zou, Chuan Tian, and Carlos Simmerling.
\newblock Blinded prediction of protein-ligand binding affinity using amber
  thermodynamic integration for the 2018 {D3R} grand challenge 4.
\newblock {\em ChemArxiv}, 6 2019.

\bibitem{wang2011replica}
Lingle Wang, Richard~A Friesner, and BJ~Berne.
\newblock Replica exchange with solute scaling: a more efficient version of
  replica exchange with solute tempering (rest2).
\newblock {\em J. Phys. Chem. B}, 115(30):9431--9438, 2011.

\bibitem{hauser2018predicting}
Kevin Hauser, Christopher Negron, Steven~K Albanese, Soumya Ray, Thomas
  Steinbrecher, Robert Abel, John~D Chodera, and Lingle Wang.
\newblock Predicting resistance of clinical abl mutations to targeted kinase
  inhibitors using alchemical free-energy calculations.
\newblock {\em Communications biology}, 1(1):70, 2018.

\bibitem{Jorgensen:Buckner:Boudon:Rives:88}
W.~L. Jorgensen, J.~K. Buckner, S.~Boudon, and J.~Tirado-Rives.
\newblock Efficient computation of absolute free energies of binding by
  computer simulations. application to the methane dimer in water.
\newblock {\em J. Chem. Phys.}, 89:3742, 1988.

\bibitem{Gilson:Given:Bush:McCammon:97}
M.~K. Gilson, J.~A. Given, B.~L. Bush, and J.~A. McCammon.
\newblock The statistical-thermodynamic basis for computation of binding
  affinities: A critical review.
\newblock {\em Biophys. J.}, 72:1047--1069, 1997.

\bibitem{Deng2009}
Yuqing Deng and Beno\^{i}t Roux.
\newblock Computations of standard binding free energies with molecular
  dynamics simulations.
\newblock {\em J. Phys. Chem. B}, 113:2234--2246, 2009.

\bibitem{Chodera:Mobley:cosb2011}
John~D. Chodera, David~L. Mobley, Michael~R. Shirts, Richard~W. Dixon, Kim
  Branson, and Vijay~S. Pande.
\newblock Alchemical free energy methods for drug discovery: Progress and
  challenges.
\newblock {\em Curr. Opin. Struct. Biol.}, 21:150--160, 2011.

\bibitem{kilburg2018assessment}
Denise Kilburg and Emilio Gallicchio.
\newblock Assessment of a single decoupling alchemical approach for the
  calculation of the absolute binding free energies of protein-peptide
  complexes.
\newblock {\em Frontiers in Molecular Biosciences}, 5:22, 2018.

\bibitem{mobley2017predicting}
David~L Mobley and Michael~K Gilson.
\newblock Predicting binding free energies: frontiers and benchmarks.
\newblock {\em Ann. Rev. Bioph.}, 46:531--558, 2017.

\bibitem{bfzhang2016rnaseh}
Baofeng Zhang, Michael~P. DErasmo, Ryan~P. Murelli, and Emilio Gallicchio.
\newblock Free energy-based virtual screening and optimization of {RNase H}
  inhibitors of {HIV-1} reverse transcriptase.
\newblock {\em ACS Omega.}, 1:435--447, 2016.

\bibitem{deng2017resolving}
Nanjie Deng, Lauren Wickstrom, Piotr Cieplak, Clement Lin, and Danzhou Yang.
\newblock Resolving the ligand-binding specificity in c-myc g-quadruplex dna:
  absolute binding free energy calculations and spr experiment.
\newblock {\em J. Phys. Chem. B}, 121(46):10484--10497, 2017.

\bibitem{kilburg2018analytical}
Denise Kilburg and Emilio Gallicchio.
\newblock Analytical model of the free energy of alchemical molecular binding.
\newblock {\em J. Chem. Theory Comput.}, 14(12):6183--6196, 2018.

\bibitem{kim2010generalized}
Jaegil Kim and John~E Straub.
\newblock Generalized simulated tempering for exploring strong phase
  transitions.
\newblock {\em J. Chem. Phys.}, 133(15):154101, 2010.

\bibitem{lu2013order}
Qing Lu, Jaegil Kim, and John~E Straub.
\newblock Order parameter free enhanced sampling of the vapor-liquid transition
  using the generalized replica exchange method.
\newblock {\em J. Chem. Phys.}, 138(10):104119, 2013.

\bibitem{McCammon:Straatsma:92}
J.~A. McCammon and T.~P. Straatsma.
\newblock Computational alchemy.
\newblock {\em Annu. Rev. Phys. Chem.}, 43:407, 1992.

\bibitem{pitera2002comparison}
Jed~W Pitera and Wilfred~F van Gunsteren.
\newblock A comparison of non-bonded scaling approaches for free energy
  calculations.
\newblock {\em Molecular Simulation}, 28(1-2):45--65, 2002.

\bibitem{AlexeiV.Finkelstein2002}
Alexei~V. Finkelstein and Oleg Ptitsyn.
\newblock {\em Protein Physics}.
\newblock Academic Press, San Diego CA, 2002.

\bibitem{Zheng2007}
Weihua Zheng, Michael Andrec, Emilio Gallicchio, and Ronald~M Levy.
\newblock Simulating replica exchange simulations of protein folding with a
  kinetic network model.
\newblock {\em Proc Natl Acad Sci U S A}, 104:15340--15345, 2007.

\bibitem{yu2014order}
Tang-Qing Yu, Pei-Yang Chen, Ming Chen, Amit Samanta, Eric Vanden-Eijnden, and
  Mark Tuckerman.
\newblock Order-parameter-aided temperature-accelerated sampling for the
  exploration of crystal polymorphism and solid-liquid phase transitions.
\newblock {\em J. Chem. Phys.}, 140(21):06B603\_1, 2014.

\bibitem{gill2018binding}
Samuel~C Gill, Nathan~M Lim, Patrick~B Grinaway, Ari{\"e}n~S Rustenburg, Josh
  Fass, Gregory~A Ross, John~D Chodera, and David~L Mobley.
\newblock Binding modes of ligands using enhanced sampling (blues): Rapid
  decorrelation of ligand binding modes via nonequilibrium candidate monte
  carlo.
\newblock {\em J. Phys. Chem. B}, 122(21):5579--5598, 2018.

\bibitem{Boresch:Karplus:2003}
S~Boresch, F~Tettinger, M~Leitgeb, and M~Karplus.
\newblock Absolute binding free energies: A quantitative approach for their
  calculation.
\newblock {\em J. Phys. Chem. B}, {107}:{9535--9551}, {2003}.

\bibitem{Gallicchio2010}
Emilio Gallicchio, Mauro Lapelosa, and Ronald~M. Levy.
\newblock Binding energy distribution analysis method ({BEDAM}) for estimation
  of protein-ligand binding affinities.
\newblock {\em J. Chem. Theory Comput.}, 6:2961--2977, 2010.

\bibitem{Gallicchio:Levy:2004}
E.~Gallicchio and {R. M.} Levy.
\newblock {AGBNP}: an analytic implicit solvent model suitable for molecular
  dynamics simulations and high-resolution modeling.
\newblock {\em J. Comput. Chem.}, 25:479--499, 2004.

\bibitem{Su:Gallicchio:Levy:2007}
Y.~Su, E.~Gallicchio, K.~Das, E.~Arnold, and {R. M.} Levy.
\newblock Linear interaction energy (lie) models for ligand binding in implicit
  solvent: Theory and application to the binding of nnrtis to hiv-1 reverse
  transcriptase.
\newblock {\em J. Chem. Theory Comput.}, 3:256--277, 2007.

\bibitem{Shirts2008a}
Michael~R Shirts and John~D Chodera.
\newblock Statistically optimal analysis of samples from multiple equilibrium
  states.
\newblock {\em J. Chem. Phys.}, 129:124105, 2008.

\bibitem{Tan2012}
Zhiqiang Tan, Emilio Gallicchio, Mauro Lapelosa, and Ronald~M. Levy.
\newblock Theory of binless multi-state free energy estimation with
  applications to protein-ligand binding.
\newblock {\em J. Chem. Phys.}, 136:144102, 2012.

\bibitem{zuckerman2010statistical}
Daniel~M Zuckerman.
\newblock {\em Statistical physics of biomolecules: an introduction}.
\newblock CRC Press, 2010.

\bibitem{Kim2010b}
Jaegil Kim, Thomas Keyes, and John~E Straub.
\newblock Generalized replica exchange method.
\newblock {\em J Chem Phys}, 132:224107, 2010.

\bibitem{lu2014investigating}
Qing Lu, Jaegil Kim, James~D Farrell, David~J Wales, and John~E Straub.
\newblock Investigating the solid-liquid phase transition of water nanofilms
  using the generalized replica exchange method.
\newblock {\em J. Chem. Phys.}, 141(18):18C525, 2014.

\bibitem{Steinbrecher2011}
Thomas Steinbrecher, InSuk Joung, and David~A. Case.
\newblock Soft-core potentials in thermodynamic integration: Comparing one- and
  two-step transformations.
\newblock {\em J. Comput. Chem.}, 32:3253--3263, 2011.

\bibitem{simonson1993free}
Thomas Simonson.
\newblock Free energy of particle insertion: an exact analysis of the origin
  singularity for simple liquids.
\newblock {\em Molecular Physics}, 80(2):441--447, 1993.

\bibitem{Buelens2012}
Floris~P. Buelens and Helmut {Grubm\"{u}ller}.
\newblock Linear-scaling soft-core scheme for alchemical free energy
  calculations.
\newblock {\em J. Comput. Chem.}, 33:25--33, 2012.

\bibitem{Sugita2000}
Yuji Sugita, Akio Kitao, and Yuko Okamoto.
\newblock Multidimensional replica-exchange method for free-energy
  calculations.
\newblock {\em J. Chem. Phys.}, 113:6042--6051, 2000.

\bibitem{Felts:Harano:Gallicchio:Levy:2004}
A.~K. Felts, Y.~Harano, E.~Gallicchio, and R.~M. Levy.
\newblock Free energy surfaces of beta-hairpin and alpha-helical peptides
  generated by replica exchange molecular dynamics with the {AGBNP} implicit
  solvent model.
\newblock {\em Proteins: Struct. Funct. Bioinf.}, 56:310--321, 2004.

\bibitem{Ravindranathan:Gallicchio:Levy:2006}
K.P. Ravindranathan, E.~Gallicchio, R.~A. Friesner, A.~E. McDermott, and R.~M.
  Levy.
\newblock Conformational equilibrium of cytochrome {P450} {BM-3} complexed with
  {N}-palmitoylglycine: A replica exchange molecular dynamics study.
\newblock {\em J. Am. Chem. Soc.}, 128:5786--5791, 2006.

\bibitem{Okumura2010}
Hisashi Okumura, Emilio Gallicchio, and Ronald~M Levy.
\newblock Conformational populations of ligand-sized molecules by replica
  exchange molecular dynamics and temperature reweighting.
\newblock {\em J. Comput. Chem.}, 31:1357--1367, 2010.

\bibitem{Woods2003}
Christopher~J. Woods, Jonathan~W. Essex, and Michael~A. King.
\newblock The development of replica-exchange-based free-energy methods.
\newblock {\em J. Phys. Chem. B}, 107:13703--13710, 2003.

\bibitem{Rick2006}
Steven~W. Rick.
\newblock Increasing the efficiency of free energy calculations using parallel
  tempering and histogram reweighting.
\newblock {\em J. Chem. Theory Comput.}, 2:939--946, 2006.

\bibitem{gallicchio2015asynchronous}
Emilio Gallicchio, Junchao Xia, William~F Flynn, Baofeng Zhang, Sade
  Samlalsingh, Ahmet Mentes, and Ronald~M Levy.
\newblock Asynchronous replica exchange software for grid and heterogeneous
  computing.
\newblock {\em Computer Physics Communications}, 196:236--246, 2015.

\bibitem{chodera2011replica}
J.D. Chodera and M.R. Shirts.
\newblock Replica exchange and expanded ensemble simulations as gibbs sampling:
  Simple improvements for enhanced mixing.
\newblock {\em J. Chem. Phys.}, 135:194110, 2011.

\bibitem{merski2015homologous}
Matthew Merski, Marcus Fischer, Trent~E Balius, Oliv Eidam, and Brian~K
  Shoichet.
\newblock Homologous ligands accommodated by discrete conformations of a buried
  cavity.
\newblock {\em Proc. Natl. Acad. Sci. (USA)}, 112(16):5039--5044, 2015.

\bibitem{Kaminski:2001}
G.~A. Kaminski, R.~A. Friesner, J.~Tirado-{R}ives, and W.~L. Jorgensen.
\newblock Evaluation and reparameterization of the {OPLS-AA} force field for
  proteins via comparison with accurate quantum chemical calculations on
  peptides.
\newblock {\em J. Phys. Chem. B}, 105:6474--6487, 2001.

\bibitem{Banks:Gallicchio:Levy:2005}
{J. L.} Banks, {J. S.} Beard, Y.~Cao, {A. E.} Cho, W.~Damm, R.~Farid, {A. K.}
  Felts, {T. A.} Halgren, {D. T.} Mainz, {J. R.} Maple, R.~Murphy, {D. M.}
  Philipp, {M. P.} Repasky, {L. Y.} Zhang, {B. J.} Berne, {R. A.} Friesner,
  E.~Gallicchio, and {R. M.} Levy.
\newblock Integrated modeling program, applied chemical theory {(IMPACT)}.
\newblock {\em J. Comp. Chem.}, 26:1752--1780, 2005.

\bibitem{eastman2017openmm}
Peter Eastman, Jason Swails, John~D Chodera, Robert~T McGibbon, Yutong Zhao,
  Kyle~A Beauchamp, Lee-Ping Wang, Andrew~C Simmonett, Matthew~P Harrigan,
  Chaya~D Stern, et~al.
\newblock Openmm 7: Rapid development of high performance algorithms for
  molecular dynamics.
\newblock {\em PLoS Comp. Bio.}, 13(7):e1005659, 2017.

\bibitem{Boyce2009}
Sarah~E Boyce, David~L Mobley, Gabriel~J Rocklin, Alan~P Graves, Ken~A Dill,
  and Brian~K Shoichet.
\newblock Predicting ligand binding affinity with alchemical free energy
  methods in a polar model binding site.
\newblock {\em J. Mol. Biol.}, 394:747--763, 2009.

\bibitem{yang2004free}
Wei Yang, Ryan Bitetti-Putzer, and Martin Karplus.
\newblock Free energy simulations: use of the reverse cumulative averaging to
  determine the equilibrated region and the time required for convergence.
\newblock {\em J. Chem. Phys.}, 120(6):2618--2628, 2003.

\bibitem{chodera2016simple}
John~D Chodera.
\newblock A simple method for automated equilibration detection in molecular
  simulations.
\newblock {\em J. Chem. Theory Comput.}, 12(4):1799--1805, 2016.

\bibitem{Pierce2012}
Levi~CT Pierce, Romelia Salomon-Ferrer, Cesar Augusto F.~de Oliveira, J~Andrew
  McCammon, and Ross~C Walker.
\newblock Routine access to millisecond time scale events with accelerated
  molecular dynamics.
\newblock {\em Journal of Chemical Theory and Computation}, 8:2997--3002, 2012.

\bibitem{xia2018improving}
Junchao Xia, William Flynn, and Ronald~M Levy.
\newblock Improving prediction accuracy of binding free energies and poses of
  hiv integrase complexes using the binding energy distribution analysis method
  with flattening potentials.
\newblock {\em J. Chem. Inf. Mod.}, 58(7):1356--1371, 2018.

\bibitem{ciccotti2004blue}
Giovanni Ciccotti and Mauro Ferrario.
\newblock Blue moon approach to rare events.
\newblock {\em Molecular Simulation}, 30(11-12):787--793, 2004.

\bibitem{barducci2008well}
Alessandro Barducci, Giovanni Bussi, and Michele Parrinello.
\newblock Well-tempered metadynamics: a smoothly converging and tunable
  free-energy method.
\newblock {\em Phys. Rev. Lett.}, 100(2):020603, 2008.

\bibitem{bussi2018metadynamics}
Giovanni Bussi, Alessandro Laio, and Pratyush Tiwary.
\newblock Metadynamics: A unified framework for accelerating rare events and
  sampling thermodynamics and kinetics.
\newblock {\em Handbook of Materials Modeling: Methods: Theory and Modeling},
  pages 1--31, 2018.

\bibitem{Baron2008a}
Riccardo Baron and J.~Andrew McCammon.
\newblock ({T}hermo)dynamic role of receptor flexibility, entropy, and motional
  correlation in protein-ligand binding.
\newblock {\em ChemPhysChem}, 9:983--988, 2008.

\bibitem{pan2017quantitative}
Albert~C Pan, Huafeng Xu, Timothy Palpant, and David~E Shaw.
\newblock Quantitative characterization of the binding and unbinding of
  millimolar drug fragments with molecular dynamics simulations.
\newblock {\em J. Chem. Theory Comput.}, 13(7):3372--3377, 2017.

\bibitem{GumbelBook}
E.~J. Gumbel.
\newblock {\em Statistics of Extremes}.
\newblock Dover Publications, New York, 2012.

\bibitem{Allen:Tildesley}
M.~P. Allen and D.~J. Tildesley.
\newblock {\em Computer Simulation of Liquids}.
\newblock Oxford University Press, New York, 1993.

\end{thebibliography}

\end{document}